\pdfoutput=1
\documentclass[11pt,a4paper]{article}

\usepackage{epsfig}
\usepackage{amsmath}
\usepackage{amssymb}
\usepackage{verbatim}
\usepackage{bm}
\usepackage{dsfont}
\usepackage[footnotesize]{caption}
\usepackage{subfigure}
\usepackage{cancel}
\usepackage{cite}
\usepackage{a4wide}
\usepackage{color}
\usepackage{hyperref}
\usepackage[all]{hypcap}
\usepackage{multirow}
\usepackage{xcolor}
\usepackage{xfrac}

\usepackage{hyperref}
\hypersetup{colorlinks=true,linkcolor=blue!90!black,citecolor=green!60!black,urlcolor=blue!90!black}

\usepackage{geometry}
\renewcommand{\baselinestretch}{1.2}

\newcommand\fl{f_{\text{low}}}
\newcommand\fh{f_{\text{high}}}
\newcommand\fp{f_{\text{p}}}

\begin{document}

%%%%%%%%%%%%%%%%%%%%%%%%%%%%%%%%%%%%%%%%%%%%%%%%%%%%%%%%%%%%%%%%%%%%%%%%%%%%%%%%%%%%%%%%%%%%%%%%%%%%

\thispagestyle{empty}

\begin{flushright}
CERN-TH-2021-107 \\
DESY 21-101      \\
\end{flushright}

\vskip 0.8 cm
\begin{center}
{\LARGE {\bf Stochastic gravitational-wave background\\\vspace{3mm}from metastable cosmic strings}}\\[12pt]

\bigskip
\bigskip 
{
{\bf{Wilfried Buchm\"uller$^\dagger$}\footnote{E-mail: \href{mailto:wilfried.buchmueller@desy.de}{wilfried.buchmueller@desy.de}}},
{\bf{Valerie Domcke$^{\ast,\ddag}$}\footnote{E-mail: \href{mailto:valerie.domcke@cern.ch}{valerie.domcke@cern.ch}}},
{\bf{Kai Schmitz$^{\ast}$}\footnote{E-mail: \href{mailto:kai.schmitz@cern.ch}{kai.schmitz@cern.ch}}}
\bigskip}\\[0pt]
\vspace{0.23cm}
{\it $^\dagger$ Deutsches Elektronen-Synchrotron DESY, 22607 Hamburg, Germany     \\\vspace{0.2cm}
     $^\ast$    Theoretical Physics Department, CERN, 1211 Geneva 23, Switzerland \\\vspace{0.2cm}
     $\ddag$    Institute of Physics, EPFL, CH-1015, Lausanne, Switzerland}       \\[20pt]
\bigskip
\end{center}

%%%%%%%%%%%%%%%%%%%%%%%%%%%%%%%%%%%%%%%%%%%%%%%%%%%%%%%%%%%%%%%%%%%%%%%%%%%%%%%%%%%%%%%%%%%%%%%%%%%%

\begin{abstract}
\noindent 
A metastable cosmic-string network is a generic consequence of many grand unified theories (GUTs) when combined with cosmic inflation.
Metastable cosmic strings are not topologically stable, but decay on cosmic time scales due to pair production of GUT monopoles.
This leads to a network consisting of metastable long strings on superhorizon scales as well as of string loops and segments on subhorizon scales.
We compute for the first time the complete stochastic gravitational-wave background (SGWB) arising from all these network constituents, including several technical improvements to both the derivation of the loop and segment contributions.
We find that the gravitational waves emitted by string loops provide the main contribution to the gravitational-wave spectrum in the relevant parameter space.
The resulting spectrum is consistent with the tentative signal observed by the NANOGrav and Parkes pulsar timing collaborations for a string tension of $G\mu \sim 10^{-11\ldots-7}$ and has ample discovery space for ground- and space-based detectors.
For GUT-scale string tensions, $G\mu \sim 10^{-8\ldots-7}$, metastable strings predict a SGWB in the LIGO--Virgo--KAGRA band that could be discovered in the near future.
\end{abstract}

%%%%%%%%%%%%%%%%%%%%%%%%%%%%%%%%%%%%%%%%%%%%%%%%%%%%%%%%%%%%%%%%%%%%%%%%%%%%%%%%%%%%%%%%%%%%%%%%%%%%

\newpage
{\hypersetup{linkcolor=black}\renewcommand{\baselinestretch}{1}\tableofcontents}

%%%%%%%%%%%%%%%%%%%%%%%%%%%%%%%%%%%%%%%%%%%%%%%%%%%%%%%%%%%%%%%%%%%%%%%%%%%%%%%%%%%%%%%%%%%%%%%%%%%%

\section{Introduction}

%%%%%%%%%%%%%%%%%%%%%%%%%%%%%%%%%%%%%%%%%%%%%%%%%%%%%%%%%%%%%%%%%%%%%%%%%%%%%%%%%%%%%%%%%%%%%%%%%%%%

The formation of cosmic defects is a generic feature of cosmological phase transitions~\cite{Kibble:1976sj}.
Defects such as monopoles and domain walls can easily overclose the universe and must therefore be avoided.
Cosmic strings, on the other hand, evolve towards a scaling regime in which their relative contribution to the total energy density of the Universe remains constant.
Cosmic strings have characteristic signatures in gravitational lensing, the cosmic microwave background (CMB), and the stochastic gravitational-wave background (SGWB) and are therefore a potentially very interesting messenger from the early universe (for reviews and references, see, e.g., Refs.~\cite{Vilenkin:2000jqa,Hindmarsh:2011qj}).

%%%%%%%%%%%%%%%%%%%%%%%%%%%%%%%%%%%%%%%%%%%%%%%%%%%%%%%%%%%%%%%%%%%%%%%%%%%%%%%%%%%%%%%%%%%%%%%%%%%%

During the past two decades, much progress has been made to describe the time evolution of a cosmic-string network (for a recent review and references, see \cite{Auclair:2019wcv}).
Since after an initial transient period, the characteristic width of cosmic strings is much smaller than the horizon, cosmic strings are often described by the Nambu--Goto (NG) action.
The cosmic-string network consists of ``long'' superhorizon strings and ``short'' subhorizon loops, which are formed in intercommutation events of long strings and which decay slowly by emitting gravitational radiation.
The approach to the scaling regime can be understood analytically in the velocity-dependent one-scale (VOS) model~\cite{Martins:1996jp,Martins:2000cs}, and it has also been established by large simulations of NG string networks~\cite{Ringeval:2005kr,BlancoPillado:2011dq}.

%%%%%%%%%%%%%%%%%%%%%%%%%%%%%%%%%%%%%%%%%%%%%%%%%%%%%%%%%%%%%%%%%%%%%%%%%%%%%%%%%%%%%%%%%%%%%%%%%%%%
 
Despite two decades of research, predictions of the SGWB signal from cosmic strings still have large uncertainties.
In a cosmic-string network, gravitational waves (GWs) are primarily emitted by oscillating string loops as well as in the form of GW bursts emitted by sharp features propagating on string loops, so-called cusps and kinks~\cite{Damour:2000wa,Damour:2001bk}.
To compute the SGWB signal, one has to know the number density of non-self-interacting loops per unit string length, which can be determined by NG string simulations~\cite{Ringeval:2005kr,BlancoPillado:2011dq}, as well as the average power radiated off in GWs by each loop.
Following essentially the same strategy, different groups have nevertheless obtained significantly different results~\cite{Lorenz:2010sm,Blanco-Pillado:2013qja,Blanco-Pillado:2017oxo,Auclair:2019wcv}.
The main differences concern the number density of small loops and the treatment of gravitational backreaction, which can smooth out string singularities.
Moreover, the entire picture of NG string loops decaying by gravitational radiation has been challenged by field-theoretic simulations suggesting a much faster decay of the network whose origin, however, remains mysterious~\cite{Hindmarsh:2021mnl,Auclair:2019wcv}.
In this paper, we follow the approach in Ref.~\cite{Blanco-Pillado:2017oxo}, relying on the evidence for long-string dominance in recent large simulations~\cite{Blanco-Pillado:2013qja} and assuming suppression of GW radiation from kinks after gravitational backreaction.

%%%%%%%%%%%%%%%%%%%%%%%%%%%%%%%%%%%%%%%%%%%%%%%%%%%%%%%%%%%%%%%%%%%%%%%%%%%%%%%%%%%%%%%%%%%%%%%%%%%%%

We consider cosmic strings associated with the spontaneous breaking of a local $\text{U}(1)$ symmetry embedded in a grand unified theory (GUT)~\cite{Vilenkin:2000jqa,Dvali:1994ms, Jeannerot:2003qv}, a prominent example being the breaking of $B\!-\!L$, the difference of baryon and lepton number~\cite{Buchmuller:2012wn}.
GUT-scale strings have a string tension in the range $G\mu \simeq 10^{-8}\ldots 10^{-6}$, which seems excluded by the SGWB bound set by pulsar timing array (PTA) experiments~\cite{Arzoumanian:2018saf,Kerr:2020qdo,Shannon:2015ect}, which constrain topologically stable cosmic strings to  $G\mu < 1.5 \times 10^{-11}$~\cite{Blanco-Pillado:2017rnf} for a standard loop size parameter $\alpha \sim 0.1$ (see below).
However, it was recently pointed out that this bound can be avoided for metastable cosmic strings, which opens a new window for a SGWB signal close to the current upper limit in the LIGO--Virgo--KAGRA (LVK) frequency band that is consistent with the PTA bounds~\cite{Buchmuller:2019gfy}.
Metastable cosmic strings decay by quantum tunneling into string segments connecting monopole--antimonopole pairs.
In the semiclassical approximation, the decay rate per string unit length is given by~\cite{Vilenkin:1982hm, Preskill:1992ck,Leblond:2009fq,Monin:2008mp}
\begin{equation}
\label{decayrate}
\Gamma_d = \frac{\mu}{2\pi}\,\exp\left(-\pi\kappa\right) \,,\quad \kappa = \frac{m^2}{\mu} \,,
\end{equation}
where $m$ is the monopole mass and $\mu$ is the string tension.
This suppresses the GW spectrum at low frequencies, rendering large string tensions compatible with PTA bounds.

%%%%%%%%%%%%%%%%%%%%%%%%%%%%%%%%%%%%%%%%%%%%%%%%%%%%%%%%%%%%%%%%%%%%%%%%%%%%%%%%%%%%%%%%%%%%%%%%%%%%%%%%%%%%55

This opens up a new window to explore GUT-scale physics with gravitational waves~\cite{Dror:2019syi,Buchmuller:2019gfy,Buchmuller:2020lbh,Chigusa:2020rks,Chakrabortty:2020otp,Samanta:2020cdk,King:2020hyd}, which has received considerable attention since the recent report by the NANOGrav collaboration of evidence for a stochastic common-spectrum process at nanohertz frequencies~\cite{Arzoumanian:2020vkk}, which has been interpreted as a SGWB in a large number of recent papers.
Beyond the astrophysical interpretation in terms of supermassive black-hole binaries~\cite{Middleton:2020asl}, possible cosmological interpretations include stable~\cite{Ellis:2020ena,Blasi:2020mfx,Samanta:2020cdk} as well as metastable strings~\cite{Buchmuller:2020lbh}.
In a first calculation, string tensions in the range $10^{-10} \lesssim G\mu \lesssim 10^{-6}$ were shown to be consistent with the NANOGrav data, with a monopole mass to string tension ratio of around $\sqrt{\kappa} \simeq 8$~\cite{Buchmuller:2020lbh}.
Such values can indeed be obtained in typical GUT models~\cite{Buchmuller:2021dtt}.  
Alternatively, quasi-stable strings with $\sqrt{\kappa} \gg 1$ associated with intermediate scales, which occur in symmetry breaking chains of GUT models~\cite{King:2021gmj}, may provide a fit to the NANOGrav data while predicting a suppressed SGWB contribution at LIGO scales.
Note that, also independent of grand unification, a SGWB signal from a cosmic-string network is a well-motivated signature for physics beyond the Standard Model (SM)~\cite{Cui:2018rwi,Gouttenoire:2019kij,Chang:2021afa}.

%%%%%%%%%%%%%%%%%%%%%%%%%%%%%%%%%%%%%%%%%%%%%%%%%%%%%%%%%%%%%%%%%%%%%%%%%%%%%%%%%%%%%%%%%%%%%%%%%%%%

In this paper, we will analyze the GW spectrum produced by a metastable string network in detail.
For metastable strings, both long superhorizon strings and short subhorizon loops decay into string segments.
There are then two qualitatively different possibilities.
In the first case, all the monopole magnetic flux is confined to the string, whereas in the second case, only some of the flux is confined, while the remaining flux is unconfined (for a review and references, see, for example,~\cite{Kibble:2015twa}).
The pattern of GW radiation is very different in the two cases.
In the first one, both loops and segments radiate GWs, with loops typically yielding the dominant contribution.
In the second one, only loops radiate GWs, whereas segments loose energy much more rapidly by radiating gauge quanta corresponding to the unconfined flux~\cite{Vachaspati:1986cc,Berezinsky:1997kd}.
Symmetry breaking in GUTs generically leads to monopoles with partially unconfined flux (for recent examples, see \cite{Lazarides:2019xai,Dror:2019syi,Buchmuller:2021dtt}).

%%%%%%%%%%%%%%%%%%%%%%%%%%%%%%%%%%%%%%%%%%%%%%%%%%%%%%%%%%%%%%%%%%%%%%%%%%%%%%%%%%%%%%%%%%%%%%%%%%%%

The GW spectrum of oscillating string segments connecting a monopole--antimonopole pair was first calculated by Martin and Vilenkin in a straight-string approximation~\cite{Martin:1996cp}.
Subsequently, Leblond, Shlaer, and Siemens computed the GW spectrum from bursts and the SGWB of metastable strings based on string segments~\cite{Leblond:2009fq}.
A crucial point in this analysis is the matching of an early scaling regime, where metastable strings behave like stable strings, to a decay regime, where new loops are no longer produced, at a time $t_s = 1 / \Gamma_d^{1/2}$, where the scaling regime ends.
In our work, we follow this approach, including also segments from decaying loops in the analysis.
Our analysis of the GW contribution emitted by loops largely follows the analysis in Ref.~\cite{Buchmuller:2019gfy}, adding a refined treatment of the time scale of the loop decay.

%%%%%%%%%%%%%%%%%%%%%%%%%%%%%%%%%%%%%%%%%%%%%%%%%%%%%%%%%%%%%%%%%%%%%%%%%%%%%%%%%%%%%%%%%%%%%%%%%%%%

The paper is organized as follows.
In Sec.~\ref{sec:analytical}, we outline the derivation of the GW spectrum emitted by cosmic-string loops and segments, providing simple analytical expressions for all relevant quantities for GW emission in the radiation era.
The more technical components of this analysis are deferred to the appendix.
In Sec.~\ref{sec:results}, we perform a numerical evaluation of the GW spectrum, demonstrating the detection prospects for PTAs and ground-based GW detectors.
Sec.~\ref{sec:conclusion} contains our conclusions.

%%%%%%%%%%%%%%%%%%%%%%%%%%%%%%%%%%%%%%%%%%%%%%%%%%%%%%%%%%%%%%%%%%%%%%%%%%%%%%%%%%%%%%%%%%%%%%%%%%%%

\section{Gravitational waves from string loops and segments}
\label{sec:analytical}

%%%%%%%%%%%%%%%%%%%%%%%%%%%%%%%%%%%%%%%%%%%%%%%%%%%%%%%%%%%%%%%%%%%%%%%%%%%%%%%%%%%%%%%%%%%%%%%%%%%%

The time evolution of a network of \textit{stable} cosmic strings emitting GWs has been extensively studied.
After an initial transient period, the network reaches a ``scaling regime'', where the number densities of long superhorizon strings and subhorizon loops per Hubble volume are preserved.
The long strings loose their energy mostly by loop formation, whereas the oscillating loops loose their energy by gravitational radiation.
This picture is strongly supported by analytical models as well as large numerical simulations.

%%%%%%%%%%%%%%%%%%%%%%%%%%%%%%%%%%%%%%%%%%%%%%%%%%%%%%%%%%%%%%%%%%%%%%%%%%%%%%%%%%%%%%%%%%%%%%%%%%%%

Much less is known for metastable cosmic-string networks characterized by a decay time $t_s = 1/\Gamma_d^{1/2}$.
It is expected that they behave similarly to stable networks at early times $t < t_s$.
However, they start decaying immediately after the phase transition in the course of which they are initially formed; and at $t_s$, typical string segments enter the horizon.
These segments then start oscillating under their own tension, radiating off GWs, and loop production stops.
At present, numerical simulations for metastable string networks do not exist.
We therefore follow the approach of Leblond, Shlaer, and Siemens~\cite{Leblond:2009fq} and match at the time $t_s$, which marks the end of the scaling regime, two different periods to each other:

%%%%%%%%%%%%%%%%%%%%%%%%%%%%%%%%%%%%%%%%%%%%%%%%%%%%%%%%%%%%%%%%%%%%%%%%%%%%%%%%%%%%%%%%%%%%%%%%%%%%

\bigskip
\noindent\begin{tabular}{l|ll}
Period    & Network constituents & GWs from \\\hline
$t < t_s$ & Loops, long superhorizon strings                             & Loops                     \\
$t > t_s$ & Loops, short subhorizon segments from long strings and loops & Loops and segments 
\end{tabular}
\bigskip

%%%%%%%%%%%%%%%%%%%%%%%%%%%%%%%%%%%%%%%%%%%%%%%%%%%%%%%%%%%%%%%%%%%%%%%%%%%%%%%%%%%%%%%%%%%%%%%%%%%%

The second period ends at a time $t_e$, which marks the end of GW emission, that is, the time when segments and loops have emitted all their energy in GWs and the string network disappears.
In the following, we shall briefly recall the main ingredients in the computation of the GW signal from loops and segments.
We shall also give explicit formulas for all relevant quantities in the radiation era, which is where most of the GW emission occurs for a large part of the parameter space.
For a detailed discussion of the matter era, we refer the reader to the appendix.

%%%%%%%%%%%%%%%%%%%%%%%%%%%%%%%%%%%%%%%%%%%%%%%%%%%%%%%%%%%%%%%%%%%%%%%%%%%%%%%%%%%%%%%%%%%%%%%%%%%%

Let us first consider stable cosmic strings, where the approach to scaling can be understood analytically in the velocity-dependent one-scale (VOS) model~\cite{Martins:1996jp,Martins:2000cs}.
The name of the VOS model derives from the fact that it assumes several length scales in the network to coincide up to constant factors: the inter-string separation $L = \left(\mu/\rho_\infty\right)^{1/2}$, the string correlation length, and the string curvature radius.
Here, $\mu$ is the energy density per unit of string length and $\rho_\infty$ is the energy density of the long strings.
In addition, the VOS model assumes the initial loop size at the time of formation to be controlled by the universal scale $L \propto H^{-1} \propto t$ times a constant factor.

%%%%%%%%%%%%%%%%%%%%%%%%%%%%%%%%%%%%%%%%%%%%%%%%%%%%%%%%%%%%%%%%%%%%%%%%%%%%%%%%%%%%%%%%%%%%%%%%%%%%

The energy density of long strings, $\rho_\infty$, is diluted in consequence of the Hubble expansion and depleted by loop production~\cite{Vilenkin:2000jqa},
\begin{equation}
\label{rhoinfty}
\frac{d\rho_\infty}{dt} = -2H \left(1+\bar{v}^2\right) \rho_\infty - \mu \int_0^\infty d\ell\,\ell\,f\left(\ell,t\right) \,,
\end{equation}
where $H = \dot{a}/a$ is the Hubble rate and $\bar{v} = \langle v^2_\infty \rangle^{1/2}$ denotes the root-mean-square velocity of long strings, which tends to a constant value in the scaling regime and which affects the redshift behavior of the network.
$f\left(\ell,t\right)$ is the loop production function, which gives the number density of non-self-interacting loops produced per unit time and unit string length.
The model accounts for scaling as a fixed point of differential equations for $L\left(t\right)$ and $\bar{v}\left(t\right)$; and in the radiation era, one finds $L\left(t\right) \rightarrow \xi t$, $\xi = 0.271$ and $\bar{v}\left(t\right) \rightarrow 0.662$.
Using Eq.~\eqref{rhoinfty}, the loop production function then takes the form~\cite{Martins:2000cs},
\begin{equation}
f\left(\ell,t\right) = \frac{A}{\alpha t^4}\,\delta\left(\ell-\alpha t\right) \,, \quad A \propto \frac{\tilde{c}\bar{v}}{\xi^3} \,,
\end{equation}
where the constant $\tilde{c}$ parametrizes the efficiency of loop-chopping.
As discussed in the appendix, this loop production function yields the loop number density
\begin{equation}
\label{vos}
\overset{\circ}{n}\left(\ell,t\right) = \frac{A}{\alpha}\,\frac{\left(\alpha+\Gamma G\mu\right)^{3/2}}{t^{3/2}\left(\ell + \Gamma G\mu t\right)^{5/2}} \,\Theta\left(\alpha t - \ell\right) 
\end{equation}
during radiation domination.
Here, $\Gamma G\mu^2$ is the total power radiated by a loop, $G$ is Newton's constant, and $A = 0.54$ is obtained from a fit to numerical simulations.

%%%%%%%%%%%%%%%%%%%%%%%%%%%%%%%%%%%%%%%%%%%%%%%%%%%%%%%%%%%%%%%%%%%%%%%%%%%%%%%%%%%%%%%%%%%%%%%%%%%%

An important quantity is the ratio of the energy densities in loops and long strings.
Using $\rho_\infty = \mu/L^2 \rightarrow \mu/\left(t\xi\right)^2$, Eq.~\eqref{vos} yields for $\alpha \gg \Gamma G\mu$,
\begin{equation}
\frac{\mu}{\rho_\infty} \int_0^\infty d\ell\,\ell\, \overset{\circ}{n}\left(\ell,t\right) = \frac{4\sqrt{\alpha}\,A\,\xi^2}{3\left(\Gamma G\mu\right)^{1/2}} \sim 10 \left(\frac{50}{\Gamma}\right)^{1/2}\left(\frac{10^{-7}}{G\mu}\right)^{1/2} \,,
\end{equation}
which increases with decreasing radiated power $\Gamma$ as well as with decreasing string tension~$\mu$.

%%%%%%%%%%%%%%%%%%%%%%%%%%%%%%%%%%%%%%%%%%%%%%%%%%%%%%%%%%%%%%%%%%%%%%%%%%%%%%%%%%%%%%%%%%%%%%%%%%%%

Large numerical simulations of cosmic-string networks have led to the number density of the Blanco-Pillado--Olum--Shlaer (BOS) model~\cite{Blanco-Pillado:2013qja}, which is similar to Eq.~\eqref{vos}.
Compared to the VOS model, also the parameter $\alpha$ is determined as $\alpha \simeq 0.1$.
The BOS number density describing the loop population in the radiation era is given by~\cite{Auclair:2019wcv}
\begin{equation}
\label{bos}
\overset{\circ}{n}\left(\ell,t\right) = \frac{B}{t^{3/2}\left(\ell+\Gamma G\mu t\right)^{5/2}}\,\Theta\left(\alpha t-\ell\right)\,\Theta\left(t_{\rm{eq}} - t\right) \,
\end{equation}
with $B\simeq 0.18$ and $t_{\rm{eq}}$ denoting the time of matter--radiation equality.
A more general discussion of loop number densities, accounting also for the loop population after matter--radiation equality, as well as corresponding references can be found in Ref.~\cite{Auclair:2019wcv}.

%%%%%%%%%%%%%%%%%%%%%%%%%%%%%%%%%%%%%%%%%%%%%%%%%%%%%%%%%%%%%%%%%%%%%%%%%%%%%%%%%%%%%%%%%%%%%%%%%%%%

Number densities for stable as well as decaying loops and string segments satisfy kinetic equations.
Their general form and solutions are described in the appendix.
The number densities for decaying loops and segments can be obtained by matching the early scaling regime to a decay regime at $t_s= 1/\Gamma_d^{1/2}$.
This procedure leads to the result for decaying loops in Eq.~\eqref{dbos1},
\begin{equation}
\label{dbos}
\overset{\circ}{n}_>\left(\ell,t\right) = \frac{B}{t^{3/2}\left(\ell + \Gamma G\mu t\right)^{5/2}}\,e^{-\Gamma_d\left[\ell\left(t-t_s\right) + \sfrac{1}{2}\,\Gamma G\mu\left(t-t_s\right)^2 \right]} \,\Theta\left(\alpha t_s- \bar{\ell}\left(t_s\right)\right) \, \Theta\left(t_{\rm{eq}}-t\right)\,, 
\end{equation}
where
\begin{equation}
\label{loft}
\bar{\ell}\left(t_s\right) = \ell + \Gamma G \mu \left(t-t_s\right) \simeq \ell + \Gamma G\mu\,t
\end{equation}
denotes the length of a loop at time $t_s$ that evolves to the length $\ell$ at time $t$ due to the emission of GWs.
The number density differs from the one for stable loops by two damping terms, which become effective for $\ell t > 1/ \Gamma_d$ and $t > \sqrt{2}\left(\Gamma G\mu\right)^{-1/2}\,t_s \equiv t_e$, respectively.
The first Heaviside theta function reflects the fact that only loops produced before $t_s$ contribute to the number density; the second theta function indicates that this expression is only valid during the radiation era.

%%%%%%%%%%%%%%%%%%%%%%%%%%%%%%%%%%%%%%%%%%%%%%%%%%%%%%%%%%%%%%%%%%%%%%%%%%%%%%%%%%%%%%%%%%%%%%%%%%%%

The number density of segments, $\tilde{n}$, can be obtained in a similar way.
It receives contributions from long strings decaying into segments as well as from loops decaying into segments, leading to the kinetic equation given in Eq.~\eqref{kintildefull}.
The former has the analytic solution~\cite{Leblond:2009fq},
\begin{equation}
\label{ntilde}
\tilde{n}^{(s)}_>\left(\ell,t\right) = C\,\frac{\Gamma_d^2}{4} \frac{\left(t+t_s\right)^2}{\sqrt{t^3t_s}} e^{-\Gamma_d\left[\ell\left(t+t_s\right) + \sfrac{1}{2}\,\tilde{\Gamma} G\mu \left(t-t_s\right)\left(t+3t_s\right)\right]} \,\Theta\left(t_{\rm eq}-t\right) \,,
\end{equation}
where $\tilde{\Gamma}$ parametrizes the GW emission of the segments and correspondingly $\tilde{t}_e = \sqrt{2}\,\big(\tilde{\Gamma}G\mu\big)^{-1/2}\,t_s$ is the time after which segments have disappeared.
Since in the scaling regime, $t < t_s$, superhorizon segments behave like stable long strings, the normalization factor $C$ can be
determined by matching the energy density of segments to the energy density $\rho_\infty$ at time $t = t_s$,
\begin{equation}
\label{matchrho}
\rho_\infty\left(t_s\right) = \frac{\mu}{t_s^2\xi^2} = \mu \int_0^\infty d\ell\,\ell\,\tilde{n}^{(s)}\left(\ell,t_s\right) = \mu\,\frac{C}{4t_s^2} \,,
\end{equation}
which yields $C = 4/\xi^2$.
For the contribution to segments from decaying loops, $\tilde{n}^{(l)}$, the kinetic equation obtains the form of a partial integro-differential equation.
We provide an exact analytical solution to this equation in terms of an infinite series in the appendix.
We, moreover, find that this exact result can be reproduced to good approximation, as far as the consequences for the GW spectrum are concerned, by multiplying the first term in the series by an overall numerical (``fudge'') factor $\sigma$,
\begin{equation}
\label{eq:ntildeloops}
\tilde{n}_>^{(l)}\left(\ell,t\right) \rightarrow \frac{\sigma}{t_s^2} \left[\ell\left(t-t_s\right) + \frac{1}{2}\,\Gamma G\mu\left(t-t_s\right)^2\right] \overset{\circ}{n}_>\left(\ell,t\right) = - \sigma\,\Gamma_d\,\frac{d}{d\Gamma_d}\,\overset{\circ}{n}_>\left(\ell,t\right) \,, 
\end{equation}
where $\sigma \simeq 5$.
The total segment number density in the radiation era for $t > t_s$ thus reads
\begin{equation}
\label{ntildefull}
\tilde{n}_>\left(\ell,t\right) = \tilde n_>^{(s)}\left(\ell,t\right) +  \tilde n_>^{(l)}\left(\ell,t\right) \,.
\end{equation}
Here and below, we follow Ref.~\cite{Leblond:2009fq} and set $\tilde \Gamma \simeq \Gamma \simeq 50$, for simplicity.
The corresponding number densities in the matter era, which enter our numerical results in Sec.~\ref{sec:results}, are derived in the appendix.

%%%%%%%%%%%%%%%%%%%%%%%%%%%%%%%%%%%%%%%%%%%%%%%%%%%%%%%%%%%%%%%%%%%%%%%%%%%%%%%%%%%%%%%%%%%%%%%%%%%%

In view of their contributions to GWs, it is interesting to compare the energy densities at $t_e$. 
From Eqs.~\eqref{dbos} and \eqref{ntildefull}, one obtains
\begin{align}
\label{energydensities}
\overset{\circ}{\rho}_>\left(t_e\right) \sim \frac{\mu}{t_e^2}\,\frac{B}                  {\left(\Gamma G\mu\right)^{1/2}} \,,\qquad
\tilde{\rho}_>^{(s)}\left(t_e\right)    \sim \frac{\mu}{t_e^2}\,\frac{1/\xi^2}            {\left(\Gamma G\mu\right)^{1/4}} \,,\qquad
\tilde\rho_>^{(l)}\left(t_e\right)      \sim \frac{\mu}{t_e^2}\,\frac{\alpha^{1/2}B\sigma}{\left(\Gamma G\mu\right)^{3/4}} \,.
\end{align}
For typical numbers, $B=0.18$, $\sigma =5$, $\xi = 0.271$, $\Gamma = 50$, $\alpha = 0.1$, and large tensions, $G\mu \sim 10^{-7}$, all contributions are roughly of similar size.
For smaller string tensions, the energy density stored in segments sourced by cosmic-string loops dominates over the other two contributions at $t_e$.

%%%%%%%%%%%%%%%%%%%%%%%%%%%%%%%%%%%%%%%%%%%%%%%%%%%%%%%%%%%%%%%%%%%%%%%%%%%%%%%%%%%%%%%%%%%%%%%%%%%%

\subsection{Stable loops}

%%%%%%%%%%%%%%%%%%%%%%%%%%%%%%%%%%%%%%%%%%%%%%%%%%%%%%%%%%%%%%%%%%%%%%%%%%%%%%%%%%%%%%%%%%%%%%%%%%%%

We first consider stable loops and the corresponding GW spectrum from a network in the scaling regime following the discussion in Ref.~\cite{Blanco-Pillado:2017oxo}.
The GW energy density relative to the critical density per logarithmic frequency unit at cosmic time $t$ is given by
\begin{equation}
\label{omega}
\Omega_{\rm gw}\left(t,f\right) = \frac{8\pi G}{3H^2\left(t\right)} f \rho_{\rm gw}\left(t,f\right) \,.
\end{equation}
Here, $H(t)$ is the Hubble rate and $\rho_{\rm gw}\left(t,f\right)$ is the energy density in GWs per frequency unit,
\begin{equation}
\label{powerintegral}
\rho_{\rm gw}\left(t,f\right) = \int_{t_i}^t \frac{dt'}{\left(1+z\left(t'\right)\right)^4}\,P_{\rm gw}\left(t',f'\right)\frac{\partial f'}{\partial f} \,,
\end{equation}
which is obtained from the redshifted power density in GWs integrated from some initial time $t_i$ to the time of observation $t$, with $f' = \left(1+z\left(t'\right)\right)f \equiv \left(1+z'\right)f$.
Loops of length $\ell$ oscillating in their $k$th harmonic emit GWs with frequency $f' = 2k/\ell$.
Hence, the power density per unit frequency is related to a loop number density $\overset{\circ}{n}\left(\ell,t'\right)$ and the power $G\mu^2P_k$ per unit length as%
\footnote{The sum has to terminate at some finite $k_{\text{max}}$ in order to avoid unphysical infinite energies related to the longest loops ever produced and also related to the finite width of a physical string.}
\begin{equation}
P_{\rm gw}\left(t',f'\right) = G\mu^2 \sum_{k=1}^{k_{\rm max}}\frac{\ell}{f'}\,\overset{\circ}{n}\left(\ell,t'\right)\,P_k \,.
\end{equation}
In the following, we focus on the contribution from cusps, which corresponds to
\begin{equation}
\label{cusps}
P_k = \frac{P_1}{k^{4/3}} \,,\quad P_1 = \frac{\Gamma}{\zeta\left(4/3\right)} \,, \quad \Gamma \simeq 50 \,.
\end{equation}
Integrating from $t_i$ to $t$ and changing variables, $dt' =- dz'/\left(H\left(z'\right)\left(1+z'\right)\right)$, Eq.~\eqref{powerintegral} yields
\begin{equation}
\label{rho}
\rho_{\rm gw}\left(t,f\right) = G\mu^2 \sum_{k=1}^{k_{\rm max}} C_k\left(t,f\right)P_k \,,
\end{equation}
with
\begin{equation}
\label{cn}
C_k\left(t,f\right) = \frac{2k}{f^2} \int_{t_i}^t \frac{dt'}{\left(1+z'\right)^5}\:\overset{\circ}{n}\left(\frac{2k}{f'},t'\right) = \frac{2k}{f^2} \int_{z\left(t\right)}^{z_i} \frac{dz'}{H\left(z'\right)\left(1+z'\right)^6}\:\overset{\circ}{n}\left(\frac{2k}{f'},t\left(z'\right)\right) \,.
\end{equation}
Note that the $z'$ integral depends only very weakly on the upper limit $z_i$. 

%%%%%%%%%%%%%%%%%%%%%%%%%%%%%%%%%%%%%%%%%%%%%%%%%%%%%%%%%%%%%%%%%%%%%%%%%%%%%%%%%%%%%%%%%%%%%%%%%%%%

The loop density is diluted by the expansion of the universe and sourced by the interactions of long strings, which is encoded in the loop production function $f\left(\ell,t\right)$,
\begin{equation}
\label{lnd}
\overset{\circ}{n}\left(\ell,t\right) = \int_{t_i}^t dt' \left(\frac{a\left(t'\right)}{a\left(t\right)}\right)^3 f\left(\bar{\ell}\left(t'\right),t'\right) \,,
\end{equation}
where $a\left(t\right)$ is the scale factor and where the function $\bar{\ell}$ has been introduced in Eq.~\eqref{loft}, $\bar{\ell}\left(t'\right) = \ell + \Gamma G\mu \left(t-t'\right)$.
In the VOS model, and approximately also in the BOS model, strings are formed with a fixed fraction $\alpha$ of the horizon, $\bar{\ell}\left(t'\right) \approx \alpha t'$.
For a particular choice of $\ell$ and $t$, the formation time $t'$ is fixed by Eq.~\eqref{loft}, which then determines the loop number density~\eqref{lnd} via the value of the scale factor at the formation time, yielding Eqs.~\eqref{vos} and \eqref{bos}, respectively.

%%%%%%%%%%%%%%%%%%%%%%%%%%%%%%%%%%%%%%%%%%%%%%%%%%%%%%%%%%%%%%%%%%%%%%%%%%%%%%%%%%%%%%%%%%%%%%%%%%%%

In the following, we discuss the resulting GW spectrum observed today, at $t = t_0$, resulting from loops emitting GWs during the radiation era.
The computation of the GW contribution emitted during the matter era is fully analogous and can be performed numerically by inserting the corresponding loop and segment number densities derived in the appendix into Eq.~\eqref{cn}.
For pedagogical reasons, we focus on the radiation epoch in this section; however, the numerical results shown in Sec.~\ref{sec:results} will contain also the contributions from the matter era.

%%%%%%%%%%%%%%%%%%%%%%%%%%%%%%%%%%%%%%%%%%%%%%%%%%%%%%%%%%%%%%%%%%%%%%%%%%%%%%%%%%%%%%%%%%%%%%%%%%%%

During radiation domination, the Hubble rate and cosmic time are given by
\begin{equation}
H\left(z\right) = \left(1+z\right)^2 H_r \,,\quad t\left(z\right) = \frac{1}{2\left(1+z\right)^2 H_r} \,,\quad H_r = H_0\sqrt{\Omega_r} \,,
\end{equation}
with $h^2 \Omega_r = 4.15 \times 10^{-5}$, $h = 0.68$~\cite{Planck:2018vyg}, where the increase of the effective number of degrees of freedom with $z$ has been neglected.
Using the BOS loop number density in Eq.~\eqref{bos}, this yields for the coefficient functions defined in Eq.~\eqref{cn},
\begin{equation}\label{cnresult}
C_k\left(t_0,f\right) = \frac{16 B H_r^2}{3f}\left[\left(\frac{4kH_r}{f}\left(1+z_{\rm eq}\right) + \Gamma G\mu\right)^{-3/2} - \left(\frac{4kH_r}{f}\left(1+z_i\right) + \Gamma G\mu\right)^{-3/2}\right] \,,
\end{equation}
where $z_{\rm eq}$ occurs as lower integration limit in Eq.~\eqref{cn} for GWs produced in the radiation era.
From this expression, one reads off the main qualitative features of the GW spectrum today.
For%
\footnote{In the following, $\fp\left(z_{\rm cut}\right)$ will always denote the frequency that corresponds to a lower or upper cutoff $z_{\rm cut}$ in the $z'$ integral and that provides an approximate lower or upper boundary of the frequency plateau, respectively.}
\begin{equation}
\label{fhigh0}
f < f_{\rm high} \equiv f_p\left(z_i\right) \,,\quad f_p\left(z\right) \equiv \frac{4 H_r}{\Gamma G \mu} \left(1+z\right) \,,
\end{equation}
only the first of the two terms in the square bracket contributes, and the behavior of this term as a function of frequency depends on whether $f$ is larger or smaller than $f_{\rm eq} \equiv f_p\left(z_{\rm eq}\right)$.
For small frequencies, i.e., $f< f_{\rm eq}$, the GW spectrum increases as $\Omega_{\rm gw} \propto f^{3/2}$, as one reads off from Eqs.~\eqref{omega} and \eqref{cnresult}.
Starting at around $f_{\rm eq}$, the GW spectrum then begins to approach a flat plateau, with the turnover frequency $f_{\rm eq}$ being inversely proportional to $G\mu$.%
\footnote{The existence of a plateau is an inherent feature of the radiation era.
A hypothetical observer within the radiation era would drop the second theta function in Eq.~\eqref{bos} and hence replace $f_{\rm eq}$ by $4 H_r/\left(\Gamma G \mu\right)$, with $H_r = \left(\rho_r/3\right)^{1/2}/M_{\rm Pl}$.}
In between $f_{\rm eq}$ and $k_{\rm max}\,f_{\rm eq}$, the behavior of the spectrum is controlled by the sum over the harmonic oscillation modes,
\begin{equation}
\sum_{k=1}^{k_{\rm eq}} \frac{1}{k^{4/3}} = H_{k_{\rm eq}}^{(4/3)} = \zeta\left(4/3\right) - \frac{3}{k_{\rm eq}^{1/3}} \left[1 + \mathcal{O}\left(\frac{1}{k_{\rm eq}}\right)\right] \,,
\end{equation}
where $k_{\rm eq}$ is the largest integer that is smaller than $f/f_{\rm eq}$ and $H_k^{(4/3)}$ is the $k$th harmonic number of order $4/3$.
For $f_{\rm eq} < f < k_{\rm max}\,f_{\rm eq}$, the deviation of the spectrum from a flat plateau therefore decays like $f^{-1/3}$.
Once the top of the plateau is reached around $f \simeq k_{\rm max}\,f_{\rm eq}$, we have
\begin{equation}
\label{power}
\sum_{k=1}^{k_{\rm max}} P_k = \frac{H_{k_{\rm max}}^{(4/3)}}{\zeta\left(4/3\right)}\,\Gamma \simeq \Gamma \,,
\end{equation}
which, together with Eqs.~\eqref{omega}, \eqref{rho}, and \eqref{cnresult}, yields the familiar result~\cite{Auclair:2019wcv},
\begin{equation}
\label{plateau}
\Omega^{\rm plateau}_{\rm gw} \simeq \frac{128\pi}{9} B\,\Omega_r\left(\frac{G\mu}{\Gamma}\right)^{1/2} \,.
\end{equation}

%%%%%%%%%%%%%%%%%%%%%%%%%%%%%%%%%%%%%%%%%%%%%%%%%%%%%%%%%%%%%%%%%%%%%%%%%%%%%%%%%%%%%%%%%%%%%%%%%%%%

The summation over all modes also plays an important role at high frequencies.
For $f_{\rm high} < f < k_{\rm max}\,f_{\rm high}$, all modes between $k_{\rm high}$ and $k_{\rm max}$ contribute with a flat plateau to the GW spectrum, where $k_{\rm high}$ is the smallest integer that is larger than $f/f_{\rm high}$, while all modes $n < f/f_{\rm high}$ decay like $f^{-1}$ (see below).
The sum over $k$ can then be expressed in terms of the Hurwitz zeta function~\cite{Blasi:2020wpy},
\begin{align}
\sum_{k=k_{\rm high}}^{k_{\rm max}} \frac{1}{k^{4/3}} & = - \zeta\left(4/3,k_{\rm max} + 1\right) + \zeta\left(4/3,k_{\rm high}\right) \\
& = - \zeta\left(4/3,k_{\rm max} + 1\right) + \frac{3}{k_{\rm high}^{1/3}} \left[1 + \mathcal{O}\left(\frac{1}{k_{\rm high}}\right)\right] \,,
\end{align}
indicating that the spectrum decreases like $f^{-1/3}$ in the interval $f_{\rm high} < f < k_{\rm max}\,f_{\rm high}$.
At even higher frequencies, the two terms in the square bracket of Eq.~\eqref{cnresult} are always of similar size, for all values of $k$.
Expanding $C_k$ at large $f$ then shows that the spectrum only receives contributions falling off like $f^{-1}$, resulting in a total spectrum summed over all modes also falling off like $f^{-1}$.

%%%%%%%%%%%%%%%%%%%%%%%%%%%%%%%%%%%%%%%%%%%%%%%%%%%%%%%%%%%%%%%%%%%%%%%%%%%%%%%%%%%%%%%%%%%%%%%%%%%%
  
\subsection{Decaying loops}

%%%%%%%%%%%%%%%%%%%%%%%%%%%%%%%%%%%%%%%%%%%%%%%%%%%%%%%%%%%%%%%%%%%%%%%%%%%%%%%%%%%%%%%%%%%%%%%%%%%%

Loops are produced in the early scaling regime, $t < t_s$.
At later times, the produced loops shrink by emitting GWs.
Their number density $\overset{\circ}{n}_>$ satisfies a kinetic equation discussed in the appendix.
Matching $\overset{\circ}{n}_>$ at $t=t_s$ to $\overset{\circ}{n}_<$ in the scaling regime by requiring $\overset{\circ}{n}_>\left(\bar{\ell}\left(t_s\right),t_s\right) = \overset{\circ}{n}_<\left(\bar{\ell}\left(t_s\right),t_s\right)$, one obtains the number density in Eq.~\eqref{dbos1}, which corresponds to the number density in Eq.~\eqref{dbos}.

%%%%%%%%%%%%%%%%%%%%%%%%%%%%%%%%%%%%%%%%%%%%%%%%%%%%%%%%%%%%%%%%%%%%%%%%%%%%%%%%%%%%%%%%%%%%%%%%%%%%

Knowing the loop number densities, we can evaluate the coefficient functions $C_k\left(t,f\right)$, see Eq.~\eqref{cn}, which determine the GW spectrum.
For GWs generated in the radiation era, one obtains
\begin{align}
& C_k\left(t_0,f\right) = \frac{2k}{f^2} \left[\int_0^{z_s}\frac{dz'}{H\left(z'\right)\left(1+z'\right)^6}\,\overset{\circ}{n}_>\left(\frac{2k}{f'},t\left(z'\right)\right) + \int_{z_s}^{z_i} \frac{dz'}{H\left(z'\right)\left(1+z'\right)^6}\,\overset{\circ}{n}_<\left(\frac{2k}{f'},t\left(z'\right)\right)\right]\nonumber\\
& = \frac{32 B H_r^3 k}{f^2} \left[\int_{z_{\rm eq}}^{z_s} dz'\,\frac{e^{-\Gamma_d\left[\frac{k}{fH_r}\left(1+z'\right)^{-3} + \frac{\Gamma G\mu}{8H_r^2}\left(1+z'\right)^{-4}\right]}}{\left[\frac{4kH_r}{f}\left(1+z'\right) + \Gamma G\mu\right]^{5/2}} + \int_{z_s}^{z_i}dz'\,\frac{1}{\left[\frac{4kH_r}{f}(1+z') + \Gamma G\mu\right]^{5/2}}\right] \label{Cdl} \,.
\end{align}
The exponential factor yields frequency-independent and frequency-dependent cutoffs $z_e$ and $z_f$, 
\begin{equation}
\label{cutoff}
1+z_e = \left(\frac{1}{\sqrt{8}H_r}\right)^{1/2} \left(\Gamma_d\,\Gamma G\mu\right)^{1/4} \equiv \left(2H_r t_e\right)^{-1/2} \,,\quad 1+z_f = \left(\frac{\Gamma_d}{fH_r}\right)^{1/3} \,,
\end{equation}
so that the range $z' < z_m = \max\left\{z_e,z_f\right\}$ does not contribute%
\footnote{With $H_r \simeq 10^{-2} H_0$, one has $z_e \simeq \left(60/H_0\right)^{1/2} \left(\Gamma_d\,\Gamma G\mu\right)^{1/4}$, which numerically coincides with the cutoffs $z_{**}$ and $z_{\rm min}$ that were defined and employed in Ref.~\cite{Leblond:2009fq} and Ref.~\cite{Buchmuller:2019gfy}, respectively.}
to the integral for $C_k\left(t_0,f\right)$.
In this section, we will for simplicity focus on the regime $z_e > z_{\rm eq}$, ensuring that the GW production is limited to the radiation-dominated regime.
This is the case for [see Eq.~\eqref{decayrate}],
\begin{equation}
\label{eq:kappaMax}
\kappa \lesssim 81 + 0.32 \,\ln\left(G\mu\right) \,.
\end{equation}

%%%%%%%%%%%%%%%%%%%%%%%%%%%%%%%%%%%%%%%%%%%%%%%%%%%%%%%%%%%%%%%%%%%%%%%%%%%%%%%%%%%%%%%%%%%%%%%%%%%%

Cutting off the first integral in Eq.~\eqref{Cdl} at $z_m$, the coefficients $C_k$ are approximately given by 
\begin{align}
\label{cnloop}
C_k\left(t_0,f\right) & = \frac{32 B H_r^3 k}{f^2} \int_{z_m}^{z_i} dz'\,\frac{1}{\left[\frac{4kH_r}{f}\left(1+z'\right) + \Gamma G\mu\right]^{5/2}} \nonumber\\
& = \frac{16 B H_r^2}{3f}\left[\left(\frac{4kH_r}{f}\left(1+z_m\right) + \Gamma G\mu\right)^{-3/2}-\left(\frac{4kH_r}{f}\left(1+z_i\right) + \Gamma G\mu\right)^{-3/2}\right] \,.
\end{align}
Compared to Eq.~\eqref{cnresult} for stable loops, the redshift $z_{\rm eq}$ has been replaced by the cutoff $z_m$.

%%%%%%%%%%%%%%%%%%%%%%%%%%%%%%%%%%%%%%%%%%%%%%%%%%%%%%%%%%%%%%%%%%%%%%%%%%%%%%%%%%%%%%%%%%%%%%%%%%%%

Above the frequency $2\,\fl $, with
\begin{equation}
\label{flow}
\fl \equiv \fp\left(z_e\right) \sim 10^{-8}\,\textrm{Hz} \left(\frac{50}{\Gamma}\right)^{3/4} \left(\frac{10^{-7}}{G\mu}\right)^{1/2} \exp{\left(-\pi\left(\frac{\kappa}{4}-16\right)\right)} \,,
\end{equation}
one has $z_m = z_e$, and for $f > \fl$, the first term in the square brackets in Eq.~\eqref{cnloop} approaches a constant.
Hence, the GW spectrum today approximately features a plateau for $\fl < f < f_{\rm high}$.
On the other hand, at smaller frequencies, one has $z_m = z_f > z_e$, and consequently the GW spectrum falls off as $f^2$ for $f < \fl$.
For frequencies above $k_{\rm max}\,f_\text{high}$, the GW spectrum falls off again like $1/f$, where for reference Eq.~\eqref{fhigh0} can be expressed as
\begin{equation}
\label{fhigh}
f_{\rm high} \sim 10^{15}\,\textrm{Hz} \left(\frac{50}{\Gamma}\right) \left(\frac{10^{-7}}{G\mu}\right)\bigg(\frac{z_i}{10^{23}}\bigg)\: \exp{\left(-\pi\left(\frac{\kappa}{4}-16\right)\right)} \ ,
\end{equation}
with a redshift $z_i \sim 10^{23}$ corresponding to a reheating temperature of $T_{\rm rh} \sim 10^{10}\,\textrm{GeV}$.

%%%%%%%%%%%%%%%%%%%%%%%%%%%%%%%%%%%%%%%%%%%%%%%%%%%%%%%%%%%%%%%%%%%%%%%%%%%%%%%%%%%%%%%%%%%%%%%%%%%%

Note that the decaying loops have to be created before $t_s$, i.e., the argument of the first theta function in the number density in Eq.~\eqref{dbos} has to be positive.
This implies
\begin{equation}
\label{Ntheta}
k < k_{\Theta}\left(f\right) \equiv \frac{f}{4H_r}\left[\frac{2\,\alpha\, H_r}{\Gamma_d^{1/2}}\left(1+z_m\right) - \frac{\Gamma G\mu}{1+z_m}\right] \,. 
\end{equation}
For the parameters given in Eq.~\eqref{flow}, one finds $k_{\Theta}\left(f\right) \sim 100\,f/\fl$.
To obtain the GW spectrum, one has to sum over all modes.
We focus on the contribution from cusps given in Eq.~\eqref{cusps}.
From Eqs.~\eqref{omega}, \eqref{rho}, \eqref{cnloop}, and \eqref{Ntheta}, one then obtains
\begin{equation}
\label{GWdl}
\Omega_{\rm gw}\left(f\right) \simeq \frac{128\pi}{9} B\,\Omega_r \left(G\mu\right)^2 \frac{\Gamma}{\zeta(4/3)} \sum_{k=1}^{k_{\Theta}(f)} \frac{1}{k^{4/3}\left[\frac{4kH_r}{f}\left(1+z_m\right) + \Gamma G\mu\right]^{3/2}} \,.
\end{equation}
For large enough frequencies, i.e., $k_{\Theta}\left(f\right) \gg 1$, and $f > \fl$, the GW spectrum reaches a plateau as for stable loops,
\begin{equation}
\label{platdl}
\Omega^{\rm plateau}_{\rm gw}\left(f\right) \simeq \frac{128\pi}{9} B\,\Omega_r \left(\frac{G\mu}{\Gamma}\right)^{1/2} \,,
\end{equation}
where now $k_{\Theta}\left(f\right)$ states contribute to the total power $\Gamma$.

%%%%%%%%%%%%%%%%%%%%%%%%%%%%%%%%%%%%%%%%%%%%%%%%%%%%%%%%%%%%%%%%%%%%%%%%%%%%%%%%%%%%%%%%%%%%%%%%%%%%

\begin{figure}
\centering
\includegraphics[width = 0.48\textwidth]{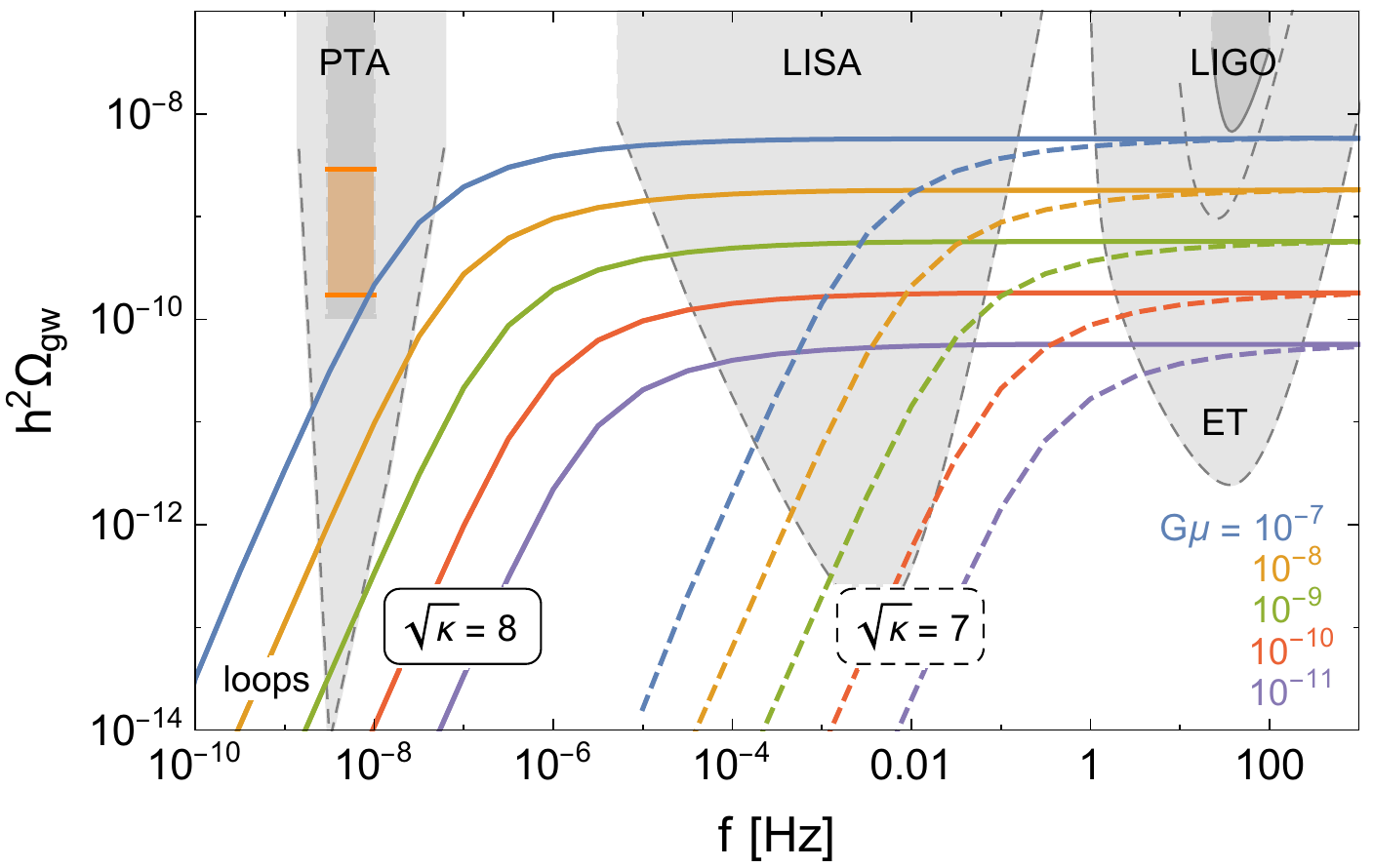}    \hfill
\includegraphics[width = 0.48\textwidth]{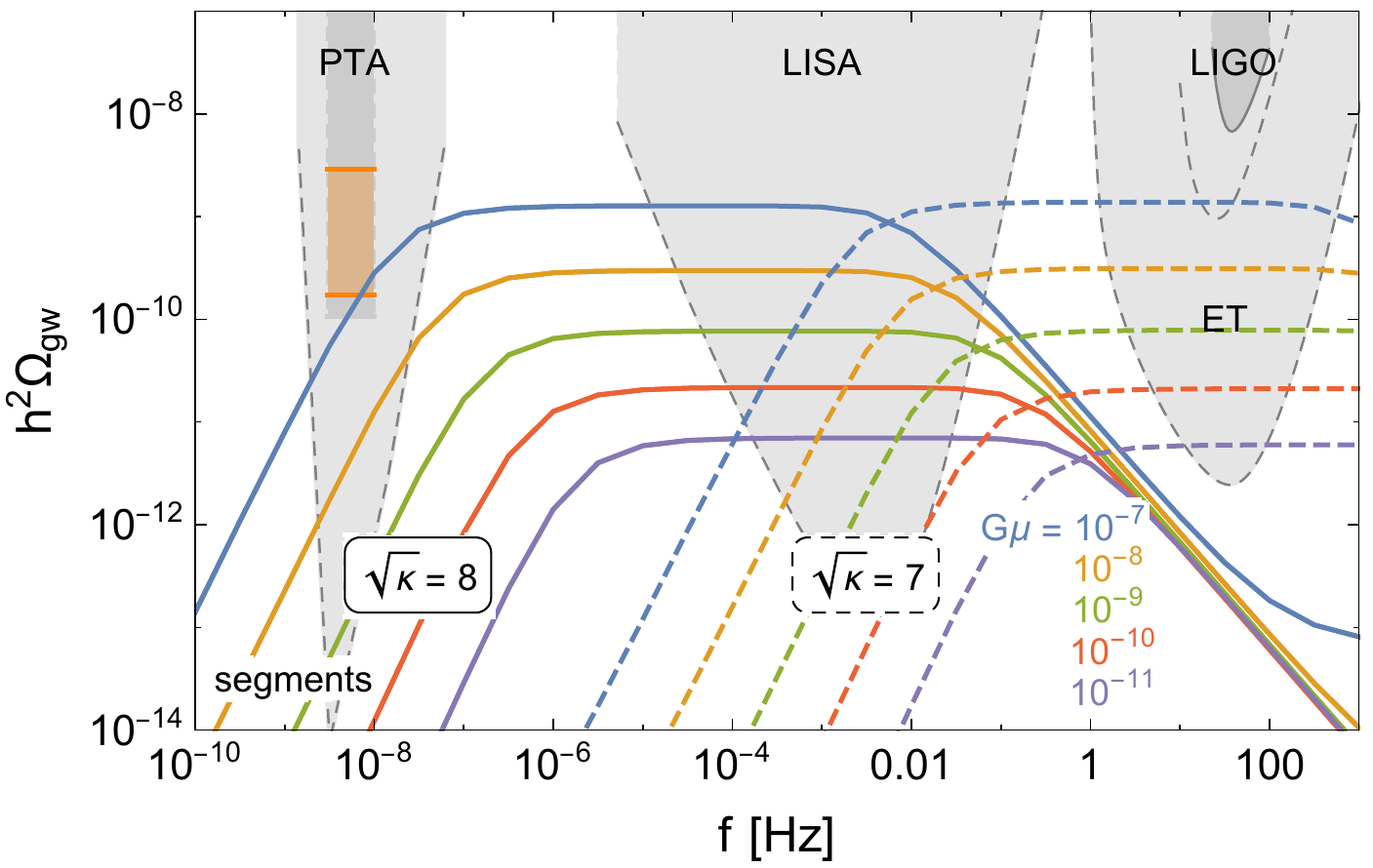}
\caption{GW spectrum from metastable cosmic-string loops (left) and segments (right) for different values of $G\mu$ and $\kappa$.
The dark gray-shaded regions indicate existing bounds from pulsar timing arrays~\cite{Shannon:2015ect} and the LIGO/VIRGO collaboration~\cite{LIGOScientific:2021iex}, the lighter shaded regions show the prospective reach of SKA~\cite{Smits:2008cf}, LISA~\cite{Audley:2017drz}, LIGO and the Einstein Telescope (ET)~\cite{ET}.
The orange shaded region indicates the region preferred by the NANOGrav hint~\cite{Arzoumanian:2020vkk}.
For simplicity, we fix the number of SM degrees of freedom to its high-temperature value in this figure, $g_* = 106.75$.}
\label{fig:spectra}
\end{figure}

The final GW spectrum sourced by loops decaying during radiation domination is shown in the left panel of Fig.~\ref{fig:spectra} for $G\mu = 10^{-11}\ldots10^{-7}$ with $\sqrt{\kappa} = 8$ (solid) and  $\sqrt{\kappa} = 7$ (dashed).
These results are obtained by inserting the loop number density in Eq.~\eqref{dbos} into Eq.~\eqref{cn}, leading to the GW energy density in Eq.~\eqref{rho}.
It suffices to take into account the loop number density at $t > t_s$, since the GW spectrum is sourced largely at $t \sim t_e \gg t_s$.
For easier comparison with the analytical results, we have fixed the number of SM degrees of freedom to its high-temperature value in this figure, $g_* = 106.75$.
The resulting agreement with the analytical expressions is very good.

%%%%%%%%%%%%%%%%%%%%%%%%%%%%%%%%%%%%%%%%%%%%%%%%%%%%%%%%%%%%%%%%%%%%%%%%%%%%%%%%%%%%%%%%%%%%%%%%%%%%

%%%%%%%%%%%%%%%%%%%%%%%%%%%%%%%%%%%%%%%%%%%%%%%%%%%%%%%%%%%%%%%%%%%%%%%%%%%%%%%%%%%%%%%%%%%%%%%%%%%%

\subsection{Decaying segments}

%%%%%%%%%%%%%%%%%%%%%%%%%%%%%%%%%%%%%%%%%%%%%%%%%%%%%%%%%%%%%%%%%%%%%%%%%%%%%%%%%%%%%%%%%%%%%%%%%%%%

At early times, $t < t_s$, the superhorizon strings decay and loose energy by chopping off loops.
As discussed in the appendix, the segment density $\tilde{n}^{(s)}_<$ sourced by long strings satisfies the kinetic equation~\cite{Leblond:2009fq}
\begin{equation}
\partial_t\, \tilde{n}^{(s)}_<\left(\ell,t\right) = -\partial_\ell \left[u\left(\ell,t\right)\tilde{n}^{(s)}_<\left(\ell,t\right)\right] - \left[3H\left(t\right) + \Gamma_d\,\ell\right] \tilde{n}^{(s)}_<\left(\ell,t\right) + 2\,\Gamma_d\,\int_{\ell} d\ell'\,\tilde{n}^{(s)}_<\left(\ell',t\right) \,,
\end{equation}
where $u\left(\ell,t\right) = 3H\left(t\right)\ell-2\ell/t$, and where the decay of segments acts as a source term for smaller segments.
The rate for producing a segment with length between $\ell$ and $\ell + d\ell'$ in the decay of a segment of length $\ell'$ is $\Gamma_d\,d\ell'$.
The segment with length $\ell$ can be chopped off at either side, hence the factor of $2$.
In the case $\ell'=2\ell$, one breaking produces two segments of length $\ell$.
The solution
\begin{equation}
\tilde{n}^{(s)}_<\left(\ell,t\right) =  C\,\Gamma_d^2\,e^{-2\,\Gamma_d\,\ell\,t}
\end{equation}
exhibits the expected scaling behaviour $\rho_{\rm cs}\left(t\right) \sim \mu \int d\ell\,\ell\,\tilde{n}^{(s)}_<\left(\ell,t\right) \sim \mu/t^2$.
The normalization constant $C= 4/\xi^2$ is determined by the scaling solution.

%%%%%%%%%%%%%%%%%%%%%%%%%%%%%%%%%%%%%%%%%%%%%%%%%%%%%%%%%%%%%%%%%%%%%%%%%%%%%%%%%%%%%%%%%%%%%%%%%%%%

At $t=t_s$, typical segments enter the horizon, loop production terminates and GW radiation begins.
Now the relevant kinetic equation reads
\begin{equation}
\partial_t\,\tilde{n}^{(s)}_>\left(\ell,t\right) = \tilde{\Gamma}\,G\mu\,\partial_\ell\,\tilde{n}^{(s)}_>\left(\ell,t\right) - \left[3H\left(t\right) + \Gamma_d\,\ell\right] \tilde{n}^{(s)}_>\left(\ell,t\right) + 2\,\Gamma_d\,\int_\ell d\ell'\,\tilde{n}^{(s)}_>\left(\ell',t\right) \,,
\end{equation}
and the solution of this integro-differential equation satisfying the initial condition $\tilde{n}^{(s)}_>\left(\bar{\ell}\left(t_s\right),t_s\right) = \tilde{n}^{(s)}_<\left(\bar{\ell}\left(t_s\right),t_s\right)$ is given by Eq.~\eqref{ntilde}.
In addition, the segment number density obtains a second contribution, $\tilde{n}^{(l)}$, from loop decays.
The exact solution of the full kinetic equation~\eqref{kintildefull} is given in the appendix.
To good approximation their contribution is described by Eq.~\eqref{eq:ntildeloops}, so that the full segment number density is then given by the sum of both contributions, see Eq.~\eqref{ntildefull}.

%%%%%%%%%%%%%%%%%%%%%%%%%%%%%%%%%%%%%%%%%%%%%%%%%%%%%%%%%%%%%%%%%%%%%%%%%%%%%%%%%%%%%%%%%%%%%%%%%%%%

Given the number density of decaying segments, we can evaluate the GW spectrum as in the previous section.
Using Eq.~\eqref{cn} but replacing the loop number density with the segment number densities given in Eqs.~\eqref{ntilde} and \eqref{eq:ntildeloops}, one obtains for the coefficient functions
\begin{align}
\label{cntilde}
\tilde{C}_k\left(t_0,f\right) & \simeq \frac{2k}{f^2} \int_0^{z_s} \frac{dz'}{H\left(z'\right)\left(1+z'\right)^6}\, \tilde{n}_>\left(\frac{2k}{f'},t\left(z'\right)\right) \\\nonumber
& \simeq \frac{2k}{f^2} \int_0^{z_s} \frac{dz'}{H\left(z'\right)\left(1+z'\right)^6} \left[-\sigma\,\Gamma_d\,\frac{\partial}{\partial \Gamma_d}\,\overset{\circ}{n}_>\left(\frac{2k}{f'},t\left(z'\right)\right) + \tilde{n}^{(s)}_>\left(\frac{2k}{f'},t\left(z'\right)\right) \right] \\\nonumber
&\simeq -\sigma\,\Gamma_d\,\frac{\partial}{\partial \Gamma_d}\,C_k\left(t_0,f\right) + \frac{2k\,\Gamma_d^2}{f^2H_r\xi^2}\int_{z_{\rm eq}}^{z_s} \frac{dz'\left(1+z_s\right)}{\left(1+z'\right)^9} \,e^{-\Gamma_d\left[\frac{k}{fH_r}\left(1+z'\right)^{-3} + \frac{\tilde{\Gamma} G\mu}{8H_r^2}\left(1+z'\right)^{-4}\right]}
\end{align}
Now the exponential yields the frequency-independent cutoff $\tilde{z}_e$, 
\begin{equation}
\label{cutofftilde}
1+\tilde{z}_e = \left(\frac{1}{\sqrt{8}H_r}\right)^{1/2} \left(\Gamma_d\,\tilde{\Gamma}G\mu\right)^{1/4} \equiv \left(2H_r \tilde{t}_e\right)^{-1/2} = \left(\frac{1}{2}\,\tilde{\Gamma} G\mu\right)^{1/4} \left(1+z_s\right) \,,
\end{equation}
and the range $z' < \tilde{z}_e$ does not contribute to the integral for $\tilde{C}_k(t,f)$.

%%%%%%%%%%%%%%%%%%%%%%%%%%%%%%%%%%%%%%%%%%%%%%%%%%%%%%%%%%%%%%%%%%%%%%%%%%%%%%%%%%%%%%%%%%%%%%%%%%%%

The power in mode $k$ of the oscillating segment is quasi-constant~\cite{Martin:1996cp},
\begin{equation} 
k\,P_k \simeq 4 \,,
\end{equation}
up to a very large maximum value determined by the Lorentz factor of the oscillating monopole, $k_{\rm max} \sim \gamma_0^2$, beyond which $P_k$ decreases like $1/k^2$.
The total power is given by
\begin{equation}
\label{Gammatilde}
\tilde{\Gamma} = \sum_{k=1}^{k_{\rm max}} P_k\sim 4 \ln{\gamma_0^2} \,,
\end{equation}
and $\tilde{\Gamma} \sim 50$ would imply $k_{\rm max} \sim 10^5$.
To obtain the GW spectrum, one has to sum over all modes and integrate over $z'$.
For $\tilde{n}^{(s)}_>$, this sum extends to $k_{\rm max}$, whereas for $\tilde{n}^{(l)}_>$ introduced above, it extends to $k_\Theta\left(f\right)$ given in Eq.~\eqref{Ntheta}.
Because of the quasi-constant behaviour of $k\,P_k$, it is convenient to perform the summation over $k$ first, approximated as an integral.
From Eqs.~\eqref{omega} and \eqref{rho} one finds for the GW spectrum, 
\begin{align}
\label{tildeOmega}
\Omega_{\rm gw}\left(f\right) \simeq \frac{32\pi \left(G\mu\right)^2}{3H_0^2f} \,\,\, \Bigg\{ & \sum_{k=1}^{k_\Theta\left(f\right)} 
\frac{8 B\sigma H_r^3 \zeta_m z_m}{\left[\frac{4kH_r}{f}\left(1+z_m\right) + \Gamma G\mu\right]^{5/2}} \\\nonumber
+ & \sum_{k=1}^{k_{\rm max}} k P_k\, \frac{\Gamma_d^2}{H_r\xi^2}\int_{\tilde{z}_e}^{z_s}\frac{dz'\left(1+z_s\right)}{\left(1+z'\right)^9}\,e^{-\frac{k\Gamma_d}{fH_r}\left(1+z'\right)^{-3}} \Bigg\} \\\nonumber
\simeq \frac{128\pi\left(G\mu\right)^2}{9} \Bigg\{ & \frac{B\sigma\zeta_m}{\left[\frac{4H_r}{f}\left(1+z_m\right)+\Gamma G\mu\right]^{3/2}} - \frac{B\sigma\zeta_m}{\left[\frac{4k_{\Theta}\left(f\right)H_r}{f}\left(1+z_m\right)+\Gamma G\mu\right]^{3/2}} \\\nonumber
+ & \frac{3\Gamma_d}{4\xi^2H_r^2} \int_{\tilde{z}_m}^{z_s}\frac{dz'\left(1+z_s\right)}{\left(1+z'\right)^6} \left[e^{-\frac{\Gamma_d}{fH_r}\left(1+z'\right)^{-3}} - e^{-\frac{\Gamma_d k_{\rm max}}{fH_r}\left(1+z'\right)^{-3}}\right]\Bigg\} \,,
\end{align}
where $\zeta_m = 1$ for $z_m = z_e$ and $\zeta_m = 4/3$ for $z_m = z_f$, and $\tilde{z}_m = \max\left(\tilde{z}_e,z_f\right)$.

%%%%%%%%%%%%%%%%%%%%%%%%%%%%%%%%%%%%%%%%%%%%%%%%%%%%%%%%%%%%%%%%%%%%%%%%%%%%%%%%%%%%%%%%%%%%%%%%%%%%

From Eq.~\eqref{tildeOmega}, the qualitative features of the GW spectrum from segments are easily understood.
For small frequencies, $f < \fl$, the first term in the bracket behaves exactly as the contribution from decaying loops in Eq.~\eqref{GWdl}, i.e., the spectrum increases as $\Omega_{\rm gw} \propto f^2$, and above $\fl$ it approaches a plateau.
However, contrary to decaying loops, the spectrum from loop segments is cut off as $k_\Theta\left(f\right)$ approaches $k_{\rm max}$ around
$\tilde{f}_{\rm high} = k_{\rm max}\,\fl$.
Above $\tilde{f}_{\rm high}$, the constant parts of the first and second terms cancel, and the spectrum falls off as $\Omega_{\rm gw} \propto 1/f$.
A similar cancellation takes place around $\tilde{f}_{\rm high}$ between the third and fourth terms arising from long-string segments, where we have again assumed $\tilde{\Gamma} = \Gamma$, i.e. $\tilde{z}_e = z_e$.
For $f < \fl$, one has $z_m = z_f$.
The $z'$ integral in Eq.~\eqref{tildeOmega} yields a factor $\left(1+z_s\right)/\left(1+z_f\right)^5$, which implies $\Omega_{\rm gw} \propto f^{5/3}$.
Finally, one finds for the height of the plateau,
\begin{equation}
\label{platseg}
\Omega_{\rm gw}\left(f\right) \simeq \frac{128 \pi}{9} \frac{\Omega_r\left(G\mu\right)^{1/2}}{\Gamma^{3/2}} \left[ B \sigma  + \frac{2^{1/4}\,12}{5 \,\xi^2} \left(\Gamma G\mu\right)^{1/4} \right] \,.
\end{equation}
Comparing the result with Eq.~\eqref{platdl}, one observes that, for $\Gamma \simeq 50$ and $\sigma \simeq 5$, the GW spectrum from loop segments is suppressed by about one order of magnitude compared to the GW spectrum from decaying loops.
The contributions from loop segments and long-string segments are about equal at $G\mu = 10^{-8}$.
Hence, the total contribution from segments to the GW spectrum is always subdominant with respect to the one from decaying loops.
A detailed comparison of the GW spectrum from decaying loops, segments from loop decays, and segments from long-string segments is given in Fig.~\ref{fig:contributions} in the appendix.
The final GW spectrum sourced by segments decaying during radiation domination is shown in the right panel of Fig.~\ref{fig:spectra} for $G\mu = 10^{-11}\ldots10^{-7}$ with $\sqrt{\kappa} = 8$ (solid) and  $\sqrt{\kappa} = 7$ (dashed).
Again we find very good agreement with our analytical expression for the GW spectrum.

%%%%%%%%%%%%%%%%%%%%%%%%%%%%%%%%%%%%%%%%%%%%%%%%%%%%%%%%%%%%%%%%%%%%%%%%%%%%%%%%%%%%%%%%%%%%%%%%%%%%
  
The extension of the plateau from the segment contribution depends on the number of contributing modes $k_{\rm max}$, which is determined by the Lorentz factor $\gamma_0^2 \sim \mu^2 \ell^2/m^2 = M_{\rm Pl}^2\,G\mu/\kappa$.
The calculation in Ref.~\cite{Martin:1996cp} obtained $\gamma_0^2 \sim 300$, corresponding to $\tilde{\Gamma} \sim 25$.
In Ref.~\cite{Leblond:2009fq}, $\tilde{\Gamma} \sim 50$ has been used, corresponding to $k_{\rm max} \sim \gamma_0^2 \sim 10^5$.
For the parameters given above, extending the plateau up to the LVK band around $100\,\textrm{Hz}$ would require $k_{\rm max} \sim 10^{10}$ with
$\tilde{\Gamma} \sim 100$.
The calculation of the radiated power has thus far been restricted to a straight string.
The extension to realistic configurations, where the two monopoles can pass each other without forming a black hole, remains a problem for future
research.%
\footnote{The GW spectrum in Eq.~\eqref{tildeOmega} can be expressed as an integral over the redshift $z'$ and the segment length $\ell$ instead of the mode number $k$~\cite{Leblond:2009fq}.
The integral over $\ell$ has the lower bound $\ell_{\rm min} \sim 1/((1+z')f)$ and the upper bound $\ell_{\rm max} \sim t_s$, i.e., the horizon at the beginning of the ``short-string period''.
Moreover, the frequency satisfies the upper bound $(1+z')f\ell < k_{\rm max} \sim \gamma_0^2 \sim \mu^2 \ell^2/m^2 = M_{\rm Pl}^2\,G\mu/\kappa$.
The integral over $z'$ is dominated by the contribution close to the lower limit $z' \sim \tilde{z}_e$.
From these inequalities, one obtains lower and upper bounds on $f$, which for large $\kappa$ read:
$\log_{10}(\fl) \sim \beta\kappa$, $\beta = -\log_{10}(e)\,\pi/4 \simeq -0.34$ and $\log_{10}(\fh) \sim \gamma\kappa$, $\gamma= 3\log_{10}(e)\,\pi/4 \simeq 1.02$.
These straight lines represent the boundaries of the plateau in figure~7 of Ref.~\cite{Leblond:2009fq}.
At $\kappa \sim 60$, the trans-Planckian upper bound is $\fh \sim 10^{70}\,\textrm{Hz}$, corresponding to $k_{\rm max} \sim 10^{78}$ and $\tilde{\Gamma} \sim 800$.}
It is expected that, above some critical mode number $k_c$, the radiated power $P_k$ falls off exponentially~\cite{Martin:1996cp} .

%%%%%%%%%%%%%%%%%%%%%%%%%%%%%%%%%%%%%%%%%%%%%%%%%%%%%%%%%%%%%%%%%%%%%%%%%%%%%%%%%%%%%%%%%%%%%%%%%%%%

\section{Detection prospects}
\label{sec:results}

%%%%%%%%%%%%%%%%%%%%%%%%%%%%%%%%%%%%%%%%%%%%%%%%%%%%%%%%%%%%%%%%%%%%%%%%%%%%%%%%%%%%%%%%%%%%%%%%%%%%
  
\begin{figure}
\centering
\includegraphics[width = 0.48 \textwidth]{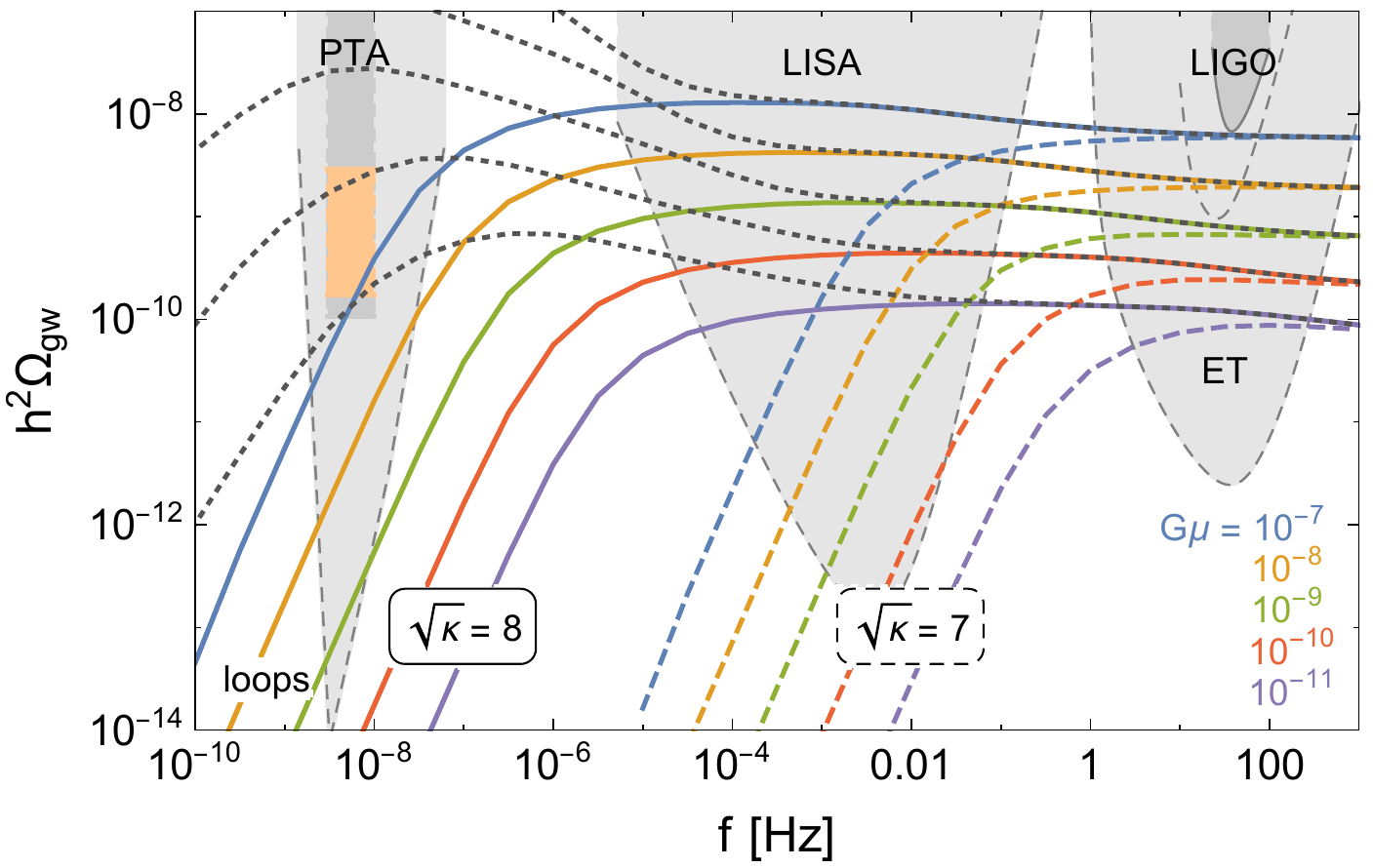} \hfill
\includegraphics[width = 0.48 \textwidth]{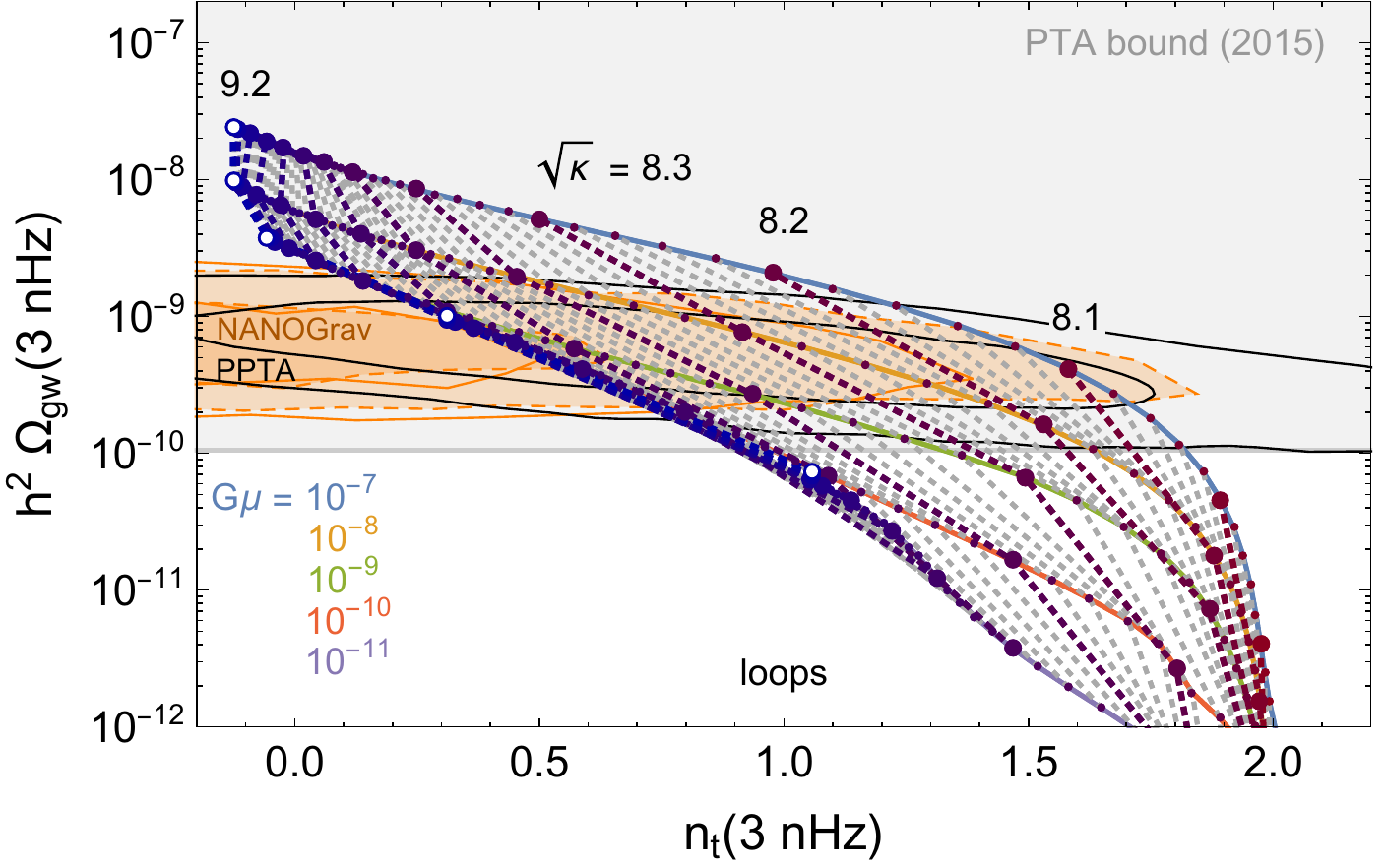}
\caption{GW spectrum from metastable cosmic strings for monopoles with unconfined fluxes.
The experimental constraints depicted in the left panel are as in Fig.~\ref{fig:spectra}.
The black dotted curves indicated the spectra obtained for topologically stable cosmic strings for the corresponding value of $G\mu$.
In the right panel, we show predictions for the frequency range of pulsar timing arrays, together with the bound from the Parkes Pulsar Timing Array (PPTA)~\cite{Shannon:2015ect} published in 2015 (gray) and the more recently reported preferred regions of NANOGrav~\cite{Arzoumanian:2020vkk} (orange) and PPTA~\cite{Goncharov:2021oub} (black).
These include contributions from the loops decaying during the matter era.}
\label{fig:summary_loops}
\end{figure}

Our final results for the GW spectrum produced by metastable cosmic strings are shown in Figs.~\ref{fig:summary_loops} to \ref{fig:summary_total}.
Starting from the analytical expressions for the loop number density in Eqs.~\eqref{dbos} and \eqref{eq:nmg} as well as (for Fig.~\ref{fig:summary_total}) the segment number density in Eqs.~\eqref{ntildefull}, \eqref{ntilde_m_s}, \eqref{eq:nSL_r}, and \eqref{eq:nSL_m}, we numerically compute the coefficient functions $C_k$ (see Eq.~\eqref{cn}) and the resulting GW spectrum (see Eqs.~\eqref{omega} and \eqref{rho}).
Here, we include the changes in the number of degrees of freedom in the SM thermal plasma, leading to deviations from the analytical prediction of a perfectly flat spectrum, particularly visible in the left panel of Fig.~\ref{fig:summary_loops}.
The summation over higher harmonics $k$ is performed up to the maximum relevant mode, as discussed in the previous section.
We vary the string tension from $G\mu = 10^{-11}$ to $10^{-7}$, covering the entire range of interest for the existing pulsar timing and ground-based interferometer experiments.
We consider values of $\sqrt{\kappa}$ all the way from quasi-stable cosmic strings, which have a life time comparable to the age of the Universe, $\sqrt{\kappa} \simeq 9$, to metastable cosmic strings with a strongly suppressed spectrum in the pulsar timing array band, $\sqrt{\kappa} \simeq 7.5$.
For this entire range of $\kappa$, the cosmic strings are stable enough to give a large signal in the LVK band.
We contrast these predictions with existing bounds, the expected sensitivity of upcoming experiments, as well as with the tentative GW signal reported by the NANOGrav collaboration~\cite{Arzoumanian:2020vkk}.

%%%%%%%%%%%%%%%%%%%%%%%%%%%%%%%%%%%%%%%%%%%%%%%%%%%%%%%%%%%%%%%%%%%%%%%%%%%%%%%%%%%%%%%%%%%%%%%%%%%%

%%%%%%%%%%%%%%%%%%%%%%%%%%%%%%%%%%%%%%%%%%%%%%%%%%%%%%%%%%%%%%%%%%%%%%%%%%%%%%%%%%%%%%%%%%%%%%%%%%%%

\subsection{Monopoles with unconfined fluxes}

%%%%%%%%%%%%%%%%%%%%%%%%%%%%%%%%%%%%%%%%%%%%%%%%%%%%%%%%%%%%%%%%%%%%%%%%%%%%%%%%%%%%%%%%%%%%%%%%%%%%

If the monopoles feature unconfined fluxes, any cosmic string segments formed from long strings or loops will rapidly decay radiating massless gauge bosons as the monopoles and antimonopoles oscillate and finally annihilate.
In this case, the resulting GW spectrum is dominated by the GW emission from cosmic string loops.
The left panel of Fig.~\ref{fig:summary_loops} shows the resulting GW spectrum for different values of the model parameters $G\mu$ and $\kappa$.
This extends the result shown in the left panel of Fig.~\ref{fig:spectra} by including the change of degrees of freedom in the SM thermal bath as well as the GW emission during the matter domination era.
\footnote{Both of these affect the cosmological expansion history, which is encoded in the Hubble parameter $H(z) = H_0 (\Omega_\Lambda + \Omega_\text{m}(1 + z)^3 + \Omega_\text{rad} {\cal G}(z) (1 + z)^4)^{1/2}$ with ${\cal G}(z) = g_*(z) g_s^{4/3}(0)/(g_*(0) g_s^{4/3}(z))$, $\Omega_\text{m} = 0.308$, $\Omega_\Lambda = 0.702$, and $g_*(z)$ ($g_s(z)$) denoting the effective number of degrees of freedom relevant for the energy (entropy) density of the SM thermal bath.}
The latter is relevant only for $t_e > t_\text{eq}$ and leads to an enhancement at low frequencies for quasi-stable strings (indicated by the dotted black curves).
We recall that the origin of this enhancement can be traced back to the scaling behaviour.
During radiation domination, the loop production and subsequent GW emission have to be efficient enough to compensate the $T^4$ decrease of the energy in the SM thermal bath.
During matter domination, scaling dictates a reduced GW emission.
In this sense, loops generated during the radiation era but surviving until the matter era radiate a disproportionate amount of energy, leading to an enhancement of the GW spectrum.

%%%%%%%%%%%%%%%%%%%%%%%%%%%%%%%%%%%%%%%%%%%%%%%%%%%%%%%%%%%%%%%%%%%%%%%%%%%%%%%%%%%%%%%%%%%%%%%%%%%%

In the right panel of Fig.~\ref{fig:summary_loops}, we perform a more detailed comparison with the existing pulsar timing results for the case of monopoles with unconfined fluxes.
Parameterizing the GW power spectrum as $\Omega_\text{gw} = \Omega_\text{gw}(f_\text{PTA}) \cdot (f/f_\text{PTA})^{n_t}$, we determine the amplitude and tilt of this power law in the range of $[2\ldots4]$~nHz around the peak sensitivity $(10~\text{yr})^{-1} \simeq 3$~nHz of current PTA experiments.
Note that, once depicted at this frequency instead of the more conventional reference frequency of $32\,\textrm{nHz} = 1/\textrm{year}$, it becomes more transparent that the NANOGrav measurement is essentially a measurement of the amplitude with the tilt still subject to a large uncertainty and largely uncorrelated with the amplitude.
We show the predictions for metastable cosmic strings for the entire range of model parameters considered, with the solid lines denoting contours of constant $G \mu$ and the dotted lines indicating contours of constant $\kappa$ (with the maximal value, $\sqrt{\kappa} = 9.2$, indicated by a white dot).
The orange region indicates the region suggested by interpreting the recent NANOGrav result as a GW signal~\cite{Arzoumanian:2020vkk}, with the solid (dashed) contours showing the $68\,\%$ and $95\,\%$ regions reported by NANOGrav when performing a fit to the first five frequency bins (when performing a fit with a broken power law).
The black solid lines show the preferred region reported by PPTA when performing a similar analysis based on the first five frequency bins~\cite{Goncharov:2021oub}.
This region is in tension with previous results from the PPTA~\cite{Shannon:2015ect} and NANOGrav~\cite{Arzoumanian:2018saf} collaborations, see Refs.~\cite{Arzoumanian:2020vkk,Hazboun:2020kzd,Goncharov:2021oub} for a discussion.

%%%%%%%%%%%%%%%%%%%%%%%%%%%%%%%%%%%%%%%%%%%%%%%%%%%%%%%%%%%%%%%%%%%%%%%%%%%%%%%%%%%%%%%%%%%%%%%%%%%%

The analysis presented here updates and largely justifies the simpler analyses performed in Refs.~\cite{Buchmuller:2019gfy,Buchmuller:2020lbh}.
The main difference is the inclusion of the decay of the cosmic string network, now encoded in the exponential function in Eq.~\eqref{dbos}.
Contrary to the previous analysis, this allows for loops of different length to decay at different times, as expected from the different probability of forming a monopole pair along the loop.
The main consequence of this is that the fall-off of the GW spectrum at small frequencies is described by an $f^2$ power law instead of $f^{3/2}$ as found in Ref.~\cite{Buchmuller:2019gfy}.
The resulting overall shift to larger values of $n_t$ for small $\kappa$ mildly reduces the overlap with the preferred NANOGrav region, implying in particular that there is now barely any overlap with the $1\sigma$ NANOGrav region, while within $2\sigma$ there is good agreement.
We note, however, that the very recent results reported by the PPTA collaboration prefer a larger spectral tilt, yielding significantly better agreement with our predictions. 
In Ref.~\cite{Buchmuller:2020lbh}, the range of string tensions $G\mu = 10^{-10} \ldots 10^{-6}$ was considered.
Given our calculation of the GW spectrum, the LIGO O3 upper bound on $\Omega_\text{gw}$~\cite{LIGOScientific:2021iex} restricts $G\mu$ to values below $2\times 10^{-7}$ (see left panel of Fig.~3).
In this paper, we therefore focus on the range of string tensions $G\mu = 10^{-11} \ldots 10^{-7}$.

%%%%%%%%%%%%%%%%%%%%%%%%%%%%%%%%%%%%%%%%%%%%%%%%%%%%%%%%%%%%%%%%%%%%%%%%%%%%%%%%%%%%%%%%%%%%%%%%%%%%
  
\begin{figure}
\centering
\includegraphics[width = 0.48 \textwidth]{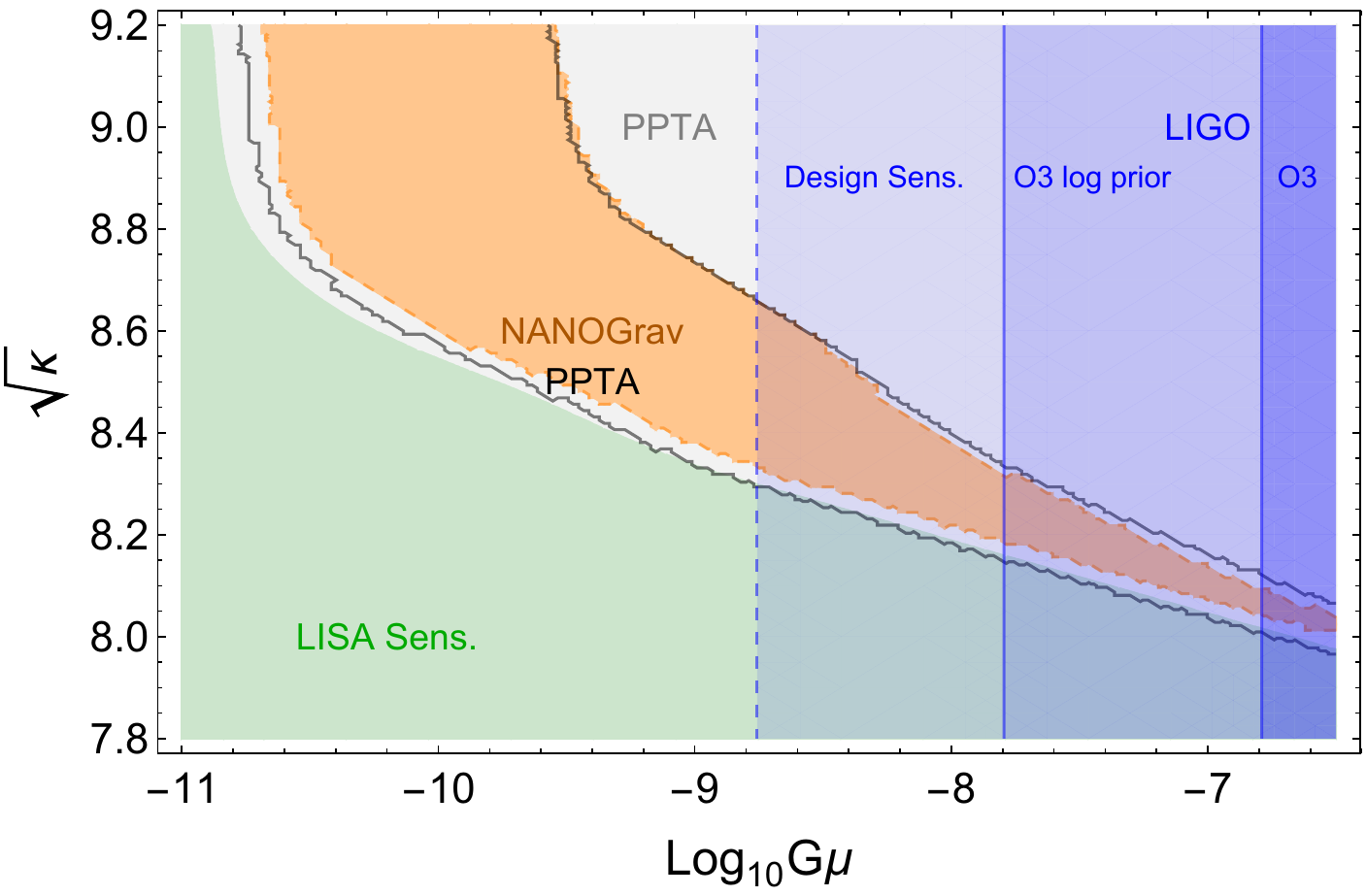} \hfill
\includegraphics[width = 0.48 \textwidth]{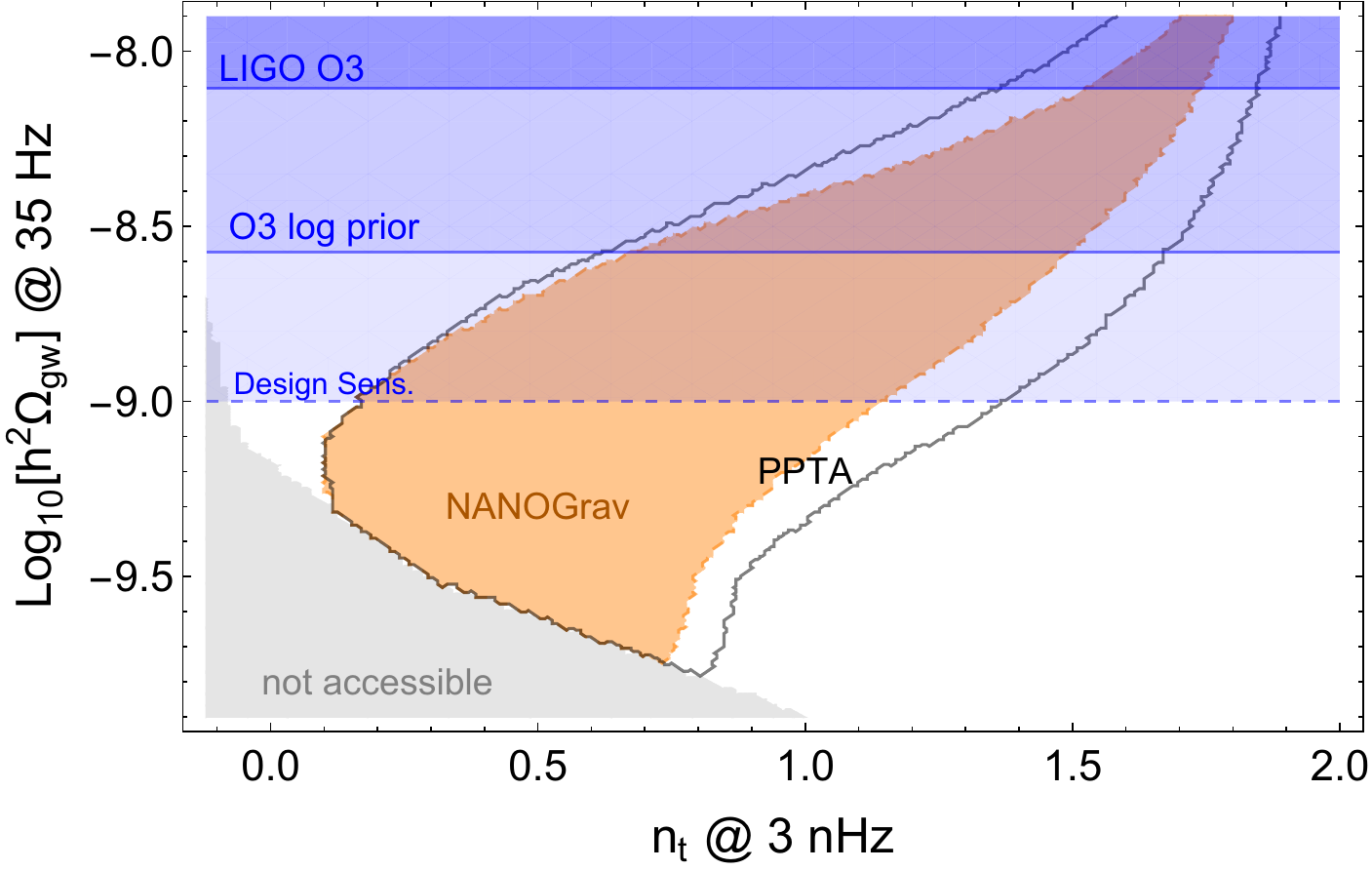}
\caption{Parameter space and GW detection prospects for monopoles with unconfined fluxes.
The orange region and the contour lines in dark gray show the tentative signal reported by NANOGrav~\cite{Arzoumanian:2020vkk} and PPTA~\cite{Goncharov:2021oub}, which lies within the exclusion region of the previous PTA bounds (grey region in left panel).
The blue shaded regions indicate bounds and prospects reported by the LIGO/VIRGO collaboration~\cite{LIGOScientific:2021iex}.
The entire parameter space shown in the left panel can be probed by LISA.
The right panel focuses on two likely future GW observables: the tilt $n_t$ of the SGWB at PTA frequencies and the amplitude at LIGO frequencies.
Metastable cosmic strings can explain a signal anywhere in this plain outside the grey shaded region.}
\label{fig:observables}
\end{figure}

%%%%%%%%%%%%%%%%%%%%%%%%%%%%%%%%%%%%%%%%%%%%%%%%%%%%%%%%%%%%%%%%%%%%year%%%%%%%%%%%%%%%%%%%%%%%%%%%%%%%%

We stress that the current significant uncertainties both in the interpretation of the PTA data as a GW signal and in the modelling of the cosmic-string network force us to take any model-to-data comparison with some grain of salt.
It is nevertheless instructive to contrast the tentative NANOGrav signal with other existing and upcoming GW observations, in particular by LIGO, see Fig.~\ref{fig:observables}.
The left panel shows the NANOGrav signal ($95\,\%\,\textrm{C.\,L.}$ region of the broken power law fit) in the model parameter plane, together with the PPTA exclusion limit~\cite{Shannon:2015ect}, the LIGO O3 bound on stochastic backgrounds~\cite{LIGOScientific:2021iex},%
\footnote{In Ref.~\cite{Abbott:2021ksc}, upper bounds on the string tension were derived for different NG models.
In a model (A) for the loop number density~\cite{Blanco-Pillado:2013qja}, the obtained GW spectrum essentially shows a plateau between the nHz and the LVK band, similar to Ref.~\cite{Blanco-Pillado:2017oxo}.
This is not the case for the loop number density of model (B)~\cite{Ringeval:2005kr}, which leads to a GW spectrum~\cite{Lorenz:2010sm} that differs from model (A) by up to four orders of magnitude.
Correspondingly, the derived bounds on the string tension are very different.
For model (A), the upper bound varies in the range $G\mu \lesssim 10^{-8}\ldots10^{-6}$, whereas for model (B), the upper bound is $G\mu \lesssim (4.0\ldots6.3) \times 10^{-15}$~\cite{Abbott:2021ksc}.
A range of two orders of magnitude for model (A) appears reasonable in view of current theoretical uncertainties.
The difference by seven orders of magnitude between models (A) and (B) is largely due to the fact that the evidence for large-loop dominance from large numerical simulations~\cite{BlancoPillado:2011dq} is not taken into account in model (B).}
and the design sensitivities of LISA and LIGO.
Note that Ref.~\cite{LIGOScientific:2021iex} quotes both a more conservative bound, $\Omega_\text{GW} < 1.7 \times 10^{-8}$ at $95\,\%\,\textrm{C.\,L.}$ (labeled ``O3'' in Fig.~\ref{fig:observables}), and a more aggressive bound, $\Omega_\text{GW} < 5.8 \times 10^{-9}$ (labeled ``O3 log prior''), depending on the choice of priors.
We conclude that, within the framework of metastable cosmic strings decaying through the production of monopoles with unconfined fluxes, the current NANOGrav data are compatible with $2 \times 10^{-11} \lesssim G\mu \lesssim 2 \times 10^{-7}$, with current (and possibly upcoming) LIGO data pushing this to lower values, towards the regime of quasi-stable cosmic strings.
We note, however, that large values of $G\mu$ are more sensitive to the formation time of the cosmic-string network and\,/\,or the reheating temperature of the Universe, which in our analysis we have taken to be at very high redshift.
Lowering this can suppress the GW spectrum at LIGO scales while leaving the predictions in the pulsar timing array band untouched.
The entire parameter space compatible with the NANOGrav signal will be finally probed by LISA.

%%%%%%%%%%%%%%%%%%%%%%%%%%%%%%%%%%%%%%%%%%%%%%%%%%%%%%%%%%%%%%%%%%%%%%%%%%%%%%%%%%%%%%%%%%%%%%%%%%%%

In the right panel of Fig.~\ref{fig:observables}, we focus on the most likely observables of the near future: the tilt measured in pulsar timing arrays (horizontal axis) and the amplitude measured in ground-based interferometers (vertical axis).
The entire white region can be reached by varying the model parameters $G\mu$ and $\kappa$, with the orange region indicating the preferred NANOGrav region.
On the contrary, a GW signal in the gray region could not be explained within this setup.

%%%%%%%%%%%%%%%%%%%%%%%%%%%%%%%%%%%%%%%%%%%%%%%%%%%%%%%%%%%%%%%%%%%%%%%%%%%%%%%%%%%%%%%%%%%%%%%%%%%%

\subsection{Monopoles with no unconfined fluxes}

%%%%%%%%%%%%%%%%%%%%%%%%%%%%%%%%%%%%%%%%%%%%%%%%%%%%%%%%%%%%%%%%%%%%%%%%%%%%%%%%%%%%%%%%%%%%%%%%%%%%

If on the contrary the monopoles do not feature any unconfined fluxes, the channel of energy loss for the cosmic-string segments is gravitational radiation.
In this case, the total GW spectrum receives an additional contribution from decaying segments, which can originate both from long strings or from loops.
The corresponding number densities are derived in the appendix and given by Eqs.~\eqref{ntildefull}, \eqref{ntilde_m_s}, \eqref{eq:nSL_r}, and \eqref{eq:nSL_m}. 

%%%%%%%%%%%%%%%%%%%%%%%%%%%%%%%%%%%%%%%%%%%%%%%%%%%%%%%%%%%%%%%%%%%%%%%%%%%%%%%%%%%%%%%%%%%%%%%%%%%%

\begin{figure}
\centering
\includegraphics[width = 0.48 \textwidth]{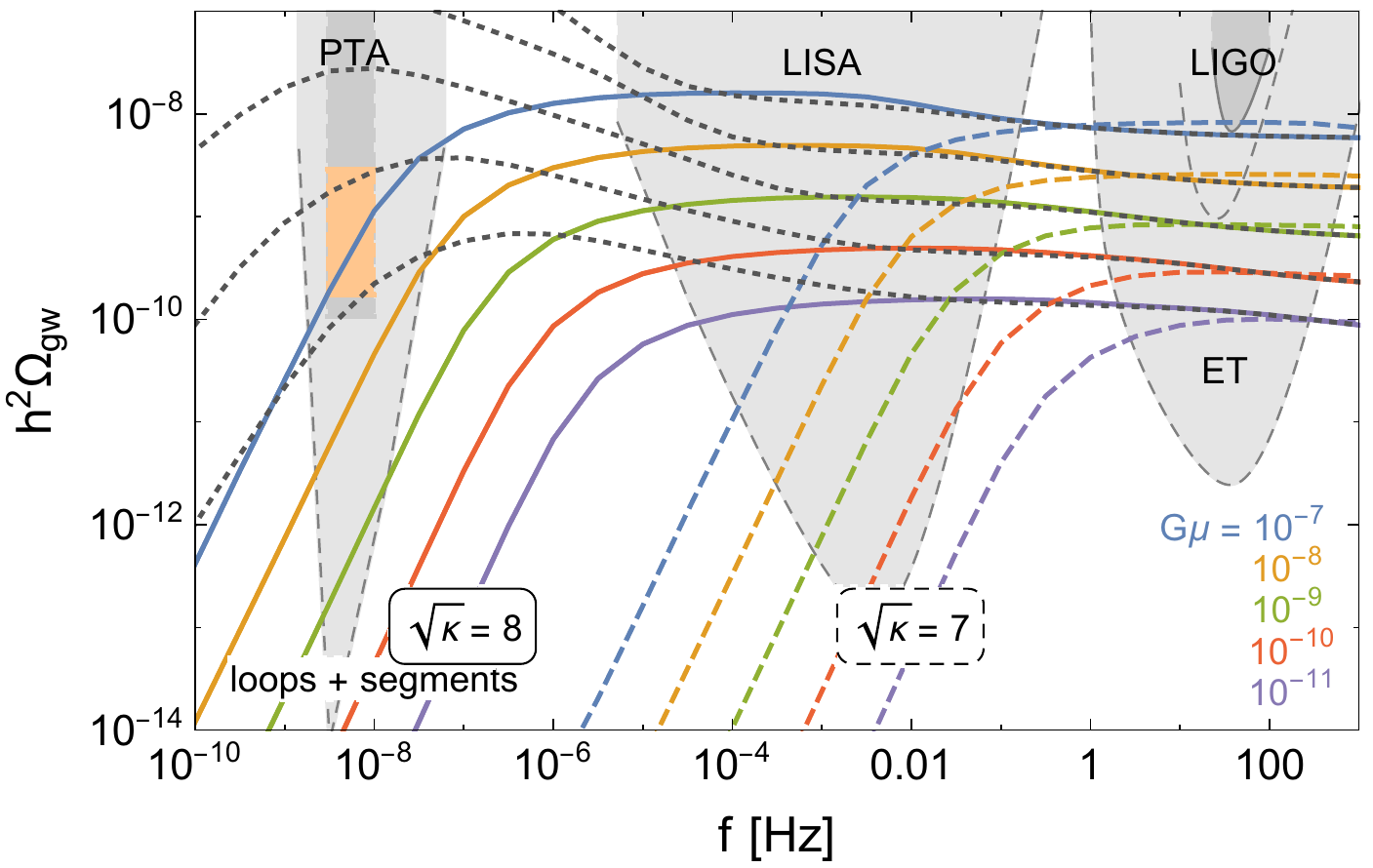} \hfill
\includegraphics[width = 0.48 \textwidth]{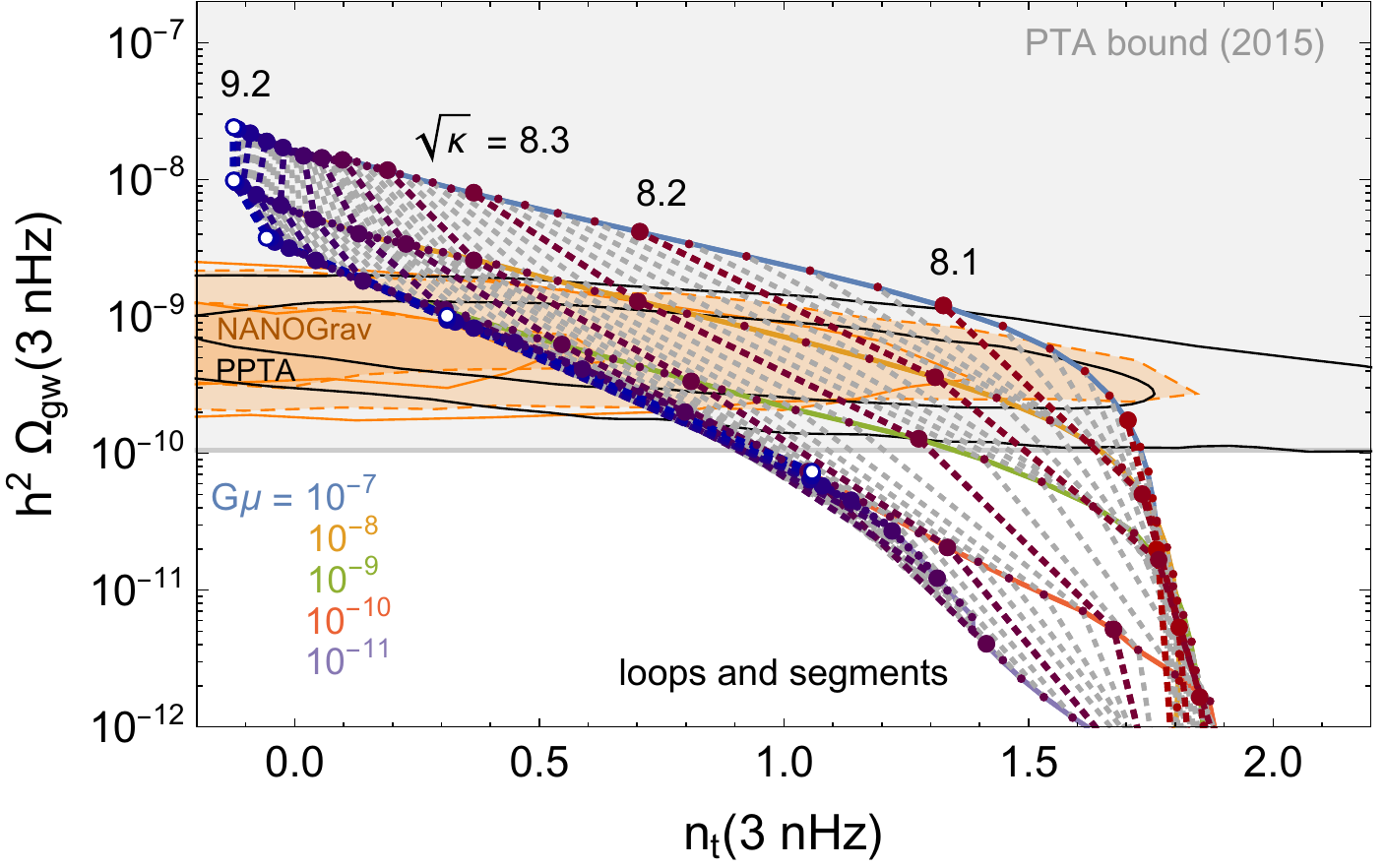}
\caption{GW spectrum from metastable cosmic strings for monopoles with no unconfined fluxes, i.e., including the additional contribution from cosmic-string segments.
Color coding as in Fig.~\ref{fig:summary_loops}.}
\label{fig:summary_total}
\end{figure}

%%%%%%%%%%%%%%%%%%%%%%%%%%%%%%%%%%%%%%%%%%%%%%%%%%%%%%%%%%%%%%%%%%%%%%%%%%%%%%%%%%%%%%%%%%%%%%%%%%%%

Fig.~\ref{fig:summary_total} shows the resulting GW spectrum as well as the predictions for pulsar timing arrays.
Comparing with Fig.~\ref{fig:summary_loops}, we note that the GW contribution from cosmic-string segments gives at most a minor correction to the GW contribution from cosmic-string loops only.
The two contributions become comparable only for large values of $G\mu$ and $f < f_{\rm low}$.
This in particular entails a flatter slope at frequencies below the onset of the plateau in radiation domination, noticeable in the right panel of Fig.~\ref{fig:summary_total} by the marginally more limited range of the tilt $n_t$ in the pulsar timing regime.
Once $\kappa$ becomes sufficiently large for the string network to survive into the matter era, the additional boost in the number density of the cosmic-string loops makes them completely dominate over the segment contribution.
For a more detailed comparison of the different contributions, see Fig.~\ref{fig:contributions} in the appendix.
In summary, we conclude that the inclusion of the GW emission from cosmic-string segments is at most a minor correction, in particular given the overall uncertainties in modelling GW emission by a cosmic-string network.
We note, however, that this conclusion is based on some model assumptions, in particular on the GW emission rate of loops and segments ($\tilde \Gamma \simeq \Gamma = 50$) and on the loop size ($\alpha = 0.1$), which may need to be revisited as our understanding of the dynamics of cosmic-string networks improves.

%%%%%%%%%%%%%%%%%%%%%%%%%%%%%%%%%%%%%%%%%%%%%%%%%%%%%%%%%%%%%%%%%%%%%%%%%%%%%%%%%%%%%%%%%%%%%%%%%%%%%

\subsection{Other observables}

%%%%%%%%%%%%%%%%%%%%%%%%%%%%%%%%%%%%%%%%%%%%%%%%%%%%%%%%%%%%%%%%%%%%%%%%%%%%%%%%%%%%%%%%%%%%%%%%%%%%%

The GW spectrum has to satisfy constraints imposed by big-bang nucleosynthesis (BBN) and the cosmic microwave background.
During BBN, the expansion rate of the universe is tightly constrained, which limits the contribution of GWs to the energy density to the contribution of about one relativistic neutrino~\cite{Caldwell:1991jj,Allen:1996vm}.
With $T_\text{BBN} \sim 0.05\,\textrm{MeV}$, one has $z_\text{BBN} \sim 5\times 10^8$ and $H(z_\text{BBN}) \sim 10^{-3}\,\textrm{Hz}$.
The contribution of one relativistic neutrino to $\Omega$ today is $\Delta\Omega_{1\nu} \simeq 7/43\,\Omega_r/(1+z_\text{eq}) \simeq 5\times 10^{-5}$.
The GW spectrum produced until $t_\text{BBN}$ has to be integrated over all subhorizon GWs present at the time of BBN.
From Eq.~\eqref{fhigh0}, one obtains for the frequency $f_\text{BBN} = f_p(z_\text{BBN}) \simeq 20\,\textrm{Hz}\,(10^{-7}/(G\mu)) \gg H(z_\text{BBN})$.
Hence, all frequencies of GWs produced before $t_\text{BBN}$ fit into the BBN Hubble horizon.
This conclusion also immediately follows from the fact that string loops and segments represent causal GW sources on subhorizon scales in a decelerating expansion background.
If the network decays after BBN, i.e. $t_e > t_\text{BBN}$, one obtains the contribution of GWs to $\Omega$ during BBN by integrating the plateau in Eq.~\eqref{plateau} from $f_\text{BBN}$ to $f_\text{high}$.
If the network decays before $t_\text{BBN}$, this integral yields an upper bound on the contribution of GWs.
In this way, one obtains
\begin{align}
\label{BBNbound}
\Delta\Omega^\text{BBN}_\text{gw} \lesssim  \int_{f_\text{BBN}}^{f_\text{high}} \frac{df}{f}\,\Omega^{\text{plateau}}_{\text{gw}} & \sim 10^{-8}\left(\frac{G\mu}{10^{-7}}\right)^{1/2} \ln{\left(\frac{f_\text{high}}{f_\text{BBN}}\right)} \\\nonumber
& \sim 10^{-8}\left(\frac{G\mu}{10^{-7}}\right)^{1/2} \left[32 + \ln{\left(\frac{T_{\text{rh}}}{10^{10}\text{GeV}} \right)}\right] \,.
\end{align} 
For the considered parameter values, $\Delta\Omega^\text{BBN}_\text{gw}$ is smaller than $\Delta\Omega_{1\nu}$ by at least three orders of magnitude.

%%%%%%%%%%%%%%%%%%%%%%%%%%%%%%%%%%%%%%%%%%%%%%%%%%%%%%%%%%%%%%%%%%%%%%%%%%%%%%%%%%%%%%%%%%%%%%%%%%%%%
    
Precision measurements of the CMB constrain cosmic-string networks in several ways.
Temperature anisotropies yield an upper bound on the tension of quasi-stable strings, $G\mu < 10^{-7}$~\cite{Ade:2015xua}, which is the largest string tension that we consider.
Other interesting observables are spectral distortions.
Current bounds are not yet very stringent, but future experiments may indeed be able to probe metastable strings~\cite{Kite:2020uix}.
In principle, also monopole annihilation from string segments could lead to interesting signatures~\cite{Bhattacharjee:1994pk,Sigl:1995kk}, which requires further investigations.

%%%%%%%%%%%%%%%%%%%%%%%%%%%%%%%%%%%%%%%%%%%%%%%%%%%%%%%%%%%%%%%%%%%%%%%%%%%%%%%%%%%%%%%%%%%%%%%%%%%%%

\section{Conclusions}
\label{sec:conclusion}

%%%%%%%%%%%%%%%%%%%%%%%%%%%%%%%%%%%%%%%%%%%%%%%%%%%%%%%%%%%%%%%%%%%%%%%%%%%%%%%%%%%%%%%%%%%%%%%%%%%%%

The formation of cosmic strings that are not topologically stable is a rather common feature in GUT models~\cite{Vilenkin:1982hm,Kibble:2015twa,Dror:2019syi}.
If the symmetry breaking step responsible for monopole production is separated from the symmetry breaking step generating cosmic strings by a phase of cosmic inflation, we generically obtain a network of metastable cosmic strings.
Their decay is triggered by pair production of monopoles along the cores of the cosmic strings.
This process is exponentially suppressed by the ratio of the monopole mass to the cosmic string tension, $\kappa = m^2/\mu$, leading to a cosmological life time.
For mass ratios in the range of $\sqrt{\kappa} \sim 7\dots8$, this leads to a strong suppression of the GW at low frequencies compared to the signal expected from topologically stable cosmic strings, while allowing for a large signal in the Hz regime.
Consequently, a large scale-invariant SGWB at LIGO frequencies would be perfectly compatible with a null detection at pulsar timing arrays.
This in particular demonstrates the significant discovery space for GWs from cosmic strings that ground-based interferometers are currently probing and which will be further significantly enlarged by the space-based interferometer LISA.
For GUT-scale string tensions,  $G\mu \sim 10^{-8} \ldots 10^{-7}$, metastable strings predict a SGWB in the LVK band that could be discovered in the very near future.

%%%%%%%%%%%%%%%%%%%%%%%%%%%%%%%%%%%%%%%%%%%%%%%%%%%%%%%%%%%%%%%%%%%%%%%%%%%%%%%%%%%%%%%%%%%%%%%%%%%%%

A significant theoretical distinction in the computation of the GW spectrum is the existence of unconfined fluxes in pair-produced GUT monopoles.
Monopoles featuring unconfined fluxes are produced as monopole--antimonopole pairs, and the resulting cosmic-string segments decay rapidly under the emission of massless gauge bosons.
On the other hand, if the monopoles do not feature any unconfined fluxes, the cosmic-string segments decay only due to GW emission, leading to an additional contribution to the GW spectrum.
In the present paper, we computed for the first time all contributions in a systematic way, allowing us to compare the different contributions in both scenarios.
In conclusion, we find that the GW contribution from cosmic-string loops is the dominant contribution in essentially the entire parameter space of interest, though if present, the contribution from cosmic-string segments has the potential to mildly influence the slope of the spectrum at PTA frequencies.

%%%%%%%%%%%%%%%%%%%%%%%%%%%%%%%%%%%%%%%%%%%%%%%%%%%%%%%%%%%%%%%%%%%%%%%%%%%%%%%%%%%%%%%%%%%%%%%%%%%%%

At the technical level, we improve the estimation of the GW spectrum from metastable cosmic-string loops first given in Ref.~\cite{Buchmuller:2019gfy} to allow for cosmic-string loops of different size decaying at different times, which changes the estimate of the low-frequency slope from $3/2$ to $2$.
For the contribution from metastable cosmic-string segments, our main finding with respect to Ref.~\cite{Leblond:2009fq} is that the frequency range of the plateau in the GW signal is limited in the ultraviolet by the number of modes contributing.
We, moreover, provide analytical formulas for the loop and segment number densities in all relevant epochs of cosmic history that take into account recent progress in the modelling topologically stable cosmic-string networks~\cite{Blanco-Pillado:2013qja}.

%%%%%%%%%%%%%%%%%%%%%%%%%%%%%%%%%%%%%%%%%%%%%%%%%%%%%%%%%%%%%%%%%%%%%%%%%%%%%%%%%%%%%%%%%%%%%%%%%%%%%

Metastable cosmic strings provide a possible explanation for the tentative SGWB signal reported by the NANOGrav~\cite{Arzoumanian:2020vkk} and PPTA~\cite{Goncharov:2021oub} collaborations.
For $2 \times 10^{-11} < G\mu <  2 \times 10^{-7}$ and $\sqrt{\kappa} > 8$, the metastable cosmic-string signal is compatible with the NANOGrav $2\sigma$ region and the PPTA $1\sigma$ region, respectively.
Upcoming, more precise determinations of the spectral tilt of this signal will be decisive in distinguishing between this explanation and other astrophysical or cosmological sources.
This, moreover, demonstrates the great potential of future multi-band GW observations involving PTAs, space-based, and ground-based interferometers. 

%%%%%%%%%%%%%%%%%%%%%%%%%%%%%%%%%%%%%%%%%%%%%%%%%%%%%%%%%%%%%%%%%%%%%%%%%%%%%%%%%%%%%%%%%%%%%%%%%%%%%

\subsubsection*{Note added}

%%%%%%%%%%%%%%%%%%%%%%%%%%%%%%%%%%%%%%%%%%%%%%%%%%%%%%%%%%%%%%%%%%%%%%%%%%%%%%%%%%%%%%%%%%%%%%%%%%%%%

Shortly after submitting this work to the arXiv, the PPTA collaboration~\cite{Goncharov:2021oub} presented an analysis of their latest data set~\cite{Kerr:2020qdo}, reporting results in agreement with NANOGrav~\cite{Arzoumanian:2020vkk} when performing a similar analysis.
For further tests and a discussion of possible interpretations of these results, see Ref.~\cite{Goncharov:2021oub}.
We included these new results in Figs.~\ref{fig:summary_loops} to \ref{fig:summary_total} in the current version, demonstrating that they fit our predictions very well.

%%%%%%%%%%%%%%%%%%%%%%%%%%%%%%%%%%%%%%%%%%%%%%%%%%%%%%%%%%%%%%%%%%%%%%%%%%%%%%%%%%%%%%%%%%%%%%%%%%%%%

\subsubsection*{Acknowledgments}

%%%%%%%%%%%%%%%%%%%%%%%%%%%%%%%%%%%%%%%%%%%%%%%%%%%%%%%%%%%%%%%%%%%%%%%%%%%%%%%%%%%%%%%%%%%%%%%%%%%%%

We thank Jose Juan Blanco-Pillado, Ryusuke Jinno, and Hitoshi Murayama for valuable discussions.
This project has received funding from the European Union's Horizon 2020 Research and Innovation Programme under grant agreement number 796961, ``AxiBAU'' (K.\,S.).

%%%%%%%%%%%%%%%%%%%%%%%%%%%%%%%%%%%%%%%%%%%%%%%%%%%%%%%%%%%%%%%%%%%%%%%%%%%%%%%%%%%%%%%%%%%%%%%%%%%%%

\appendix
\numberwithin{equation}{section}

%%%%%%%%%%%%%%%%%%%%%%%%%%%%%%%%%%%%%%%%%%%%%%%%%%%%%%%%%%%%%%%%%%%%%%%%%%%%%%%%%%%%%%%%%%%%%%%%%%%%%

\section{Kinetic equations for number densities}
\label{app:kinetic_eq}

%%%%%%%%%%%%%%%%%%%%%%%%%%%%%%%%%%%%%%%%%%%%%%%%%%%%%%%%%%%%%%%%%%%%%%%%%%%%%%%%%%%%%%%%%%%%%%%%%%%%%

The number densities of string loops and segments satisfy kinetic equations, which take various effects into account that influence the time evolution of the decaying cosmic string network.
They have the general form
\begin{equation}
\label{generalkinetic}
\partial_t\,n\left(\ell,t\right) = S\left(\ell,t\right) - \partial_\ell \left[u\left(\ell,t\right)\,n\left(\ell,t\right)\right] - \left[3H\left(t\right) + \Gamma_d\,\ell\right] n\left(\ell,t\right)\ ,
\end{equation}
where $S$ is a source term, $u$ describes the change of the string length due to Hubble stretching and energy loss of the network, $H$ is the Hubble parameter and $\Gamma_d$ is the decay rate.
Eq.~\eqref{generalkinetic} can be derived by considering the changes in $n\left(\ell,t\right)\Delta \ell$, i.e., the number density of loops whose lengths lie in the interval $\left[\ell,\ell+\Delta\ell\right]$, after some infinitesimally small time step from $t$ to $t+\Delta t$,
\begin{equation}
\label{eq:deltaldeltat}
n\left(\ell + u\,\Delta t, t+\Delta t\right)\Delta\ell' = S\left(\ell,t\right)\Delta t\,\Delta\ell + \left(\frac{a\left(t\right)}{a\left(t+\Delta t\right)}\right)^3 n\left(\ell,t\right)\Delta\ell - \Gamma_d\,\ell\,n\left(\ell,t\right)\Delta t\,\Delta\ell \,,
\end{equation}
where $\Delta\ell' = \Delta\ell + \partial_\ell u\,\Delta t\,\Delta \ell$ accounts for the change in the interval length $\Delta\ell$ during $\Delta t$.
Expanding all terms in Eq.~\eqref{eq:deltaldeltat} up to first order in $\Delta t$ and collecting all terms of order $\Delta t\,\Delta\ell$ on both sides reproduces Eq.~\eqref{generalkinetic}. 
In standard form, the partial differential equation~\eqref{generalkinetic} reads
\begin{equation}
\left[u\left(\ell,t\right) \partial_\ell + w\left(\ell,t\right) \partial_t \right] n\left(\ell,t\right) = F\left(\ell,t,n\left(\ell,t\right)\right) \,, \quad  w = 1 \,, \quad F = S - \left( 3H+\Gamma_d\,\ell + \partial_\ell u\right)n \,,
\end{equation}  
which can be solved by integrating the ordinary differential equations for the three characteristic curves $\bar{l}(t')$, $\bar{t}(t')$ and $\bar{n}(t')$ as functions of an auxiliary parameter $t' \in [t_i,t]$,
\begin{equation}
\frac{d\bar{\ell}}{dt'} = u\left(\bar{\ell},\bar{t}\right) \,, \quad \frac{d\bar{t}}{dt'} = 1 \,,\quad \frac{d\bar{n}}{dt'} = F\left(\bar{\ell},\bar{t},\bar{n}\left(\bar{\ell},\bar{t}\right)\right) \,.
\end{equation}
Imposing the boundary conditions $\bar{\ell}\left(t\right) = \ell$, $\bar{t}\left(t\right) = t$, and $\bar{n}\left(\bar{\ell}\left(t_i\right),t_i\right) = n_i\left(\bar{\ell}\left(t_i\right)\right)$, one obtains
\begin{align}
\label{solution}
n\left(\ell,t\right) & = \bar{n}\left(\bar{\ell}\left(t\right),\bar{t}\left(t\right)\right) \nonumber\\
& = \exp\left[- \int_{t_i}^t dt' \left(3H\left(\bar{t}\right)+\Gamma_d\,\bar{\ell} + \partial_{\bar{\ell}}\,u\left(\bar{\ell},\bar{t}\right)\right)\right] \times \Bigg\{ n_i\left(\bar{\ell}\left(t_i\right)\right) \nonumber \\
& \quad + \int_{t_i}^t dt' S\left(\bar{\ell},\bar{t}\right) \exp\left[\int_{t_i}^{t'}dt'' \left(3H\left(\bar{t}\right)+\Gamma_d\,\bar{\ell} + \partial_{\bar{\ell}}\,u\left(\bar{\ell},\bar{t}\right)\right)\right]\Bigg\} \nonumber\\
& = \exp\left[- \int_{t_i}^t dt' \left(\Gamma_d\,\bar{\ell}\left(t'\right) + \partial_{\bar{\ell}}\,u\left(\bar{\ell}\left(t'\right),\bar{t}\left(t'\right)\right)\right)\right] \times \Bigg\{ \left(\frac{a\left(t_i\right)}{a\left(t\right)}\right)^3 n_i\left(\bar{\ell}\left(t_i\right)\right) \nonumber \\
& \quad + \int_{t_i}^t dt' \left(\frac{a\left(t'\right)}{a\left(t\right)}\right)^3 S\left(\bar{\ell}\left(t'\right),t'\right) \exp\left[\int_{t_i}^{t'}dt'' \left(\Gamma_d\,\bar{\ell}\left(t''\right) + \partial_{\bar{\ell}}\,u\left(\bar{\ell}\left(t''\right),\bar{t}\left(t''\right)\right)\right)\right] \Bigg\} \,.
\end{align}
  
%%%%%%%%%%%%%%%%%%%%%%%%%%%%%%%%%%%%%%%%%%%%%%%%%%%%%%%%%%%%%%%%%%%%%%%%%%%%%%%%%%%%%%%%%%%%%%%%%%%%%
  
\subsection{Loops}  

%%%%%%%%%%%%%%%%%%%%%%%%%%%%%%%%%%%%%%%%%%%%%%%%%%%%%%%%%%%%%%%%%%%%%%%%%%%%%%%%%%%%%%%%%%%%%%%%%%%%%

\begin{figure}
\centering
\includegraphics[width = 0.48 \textwidth]{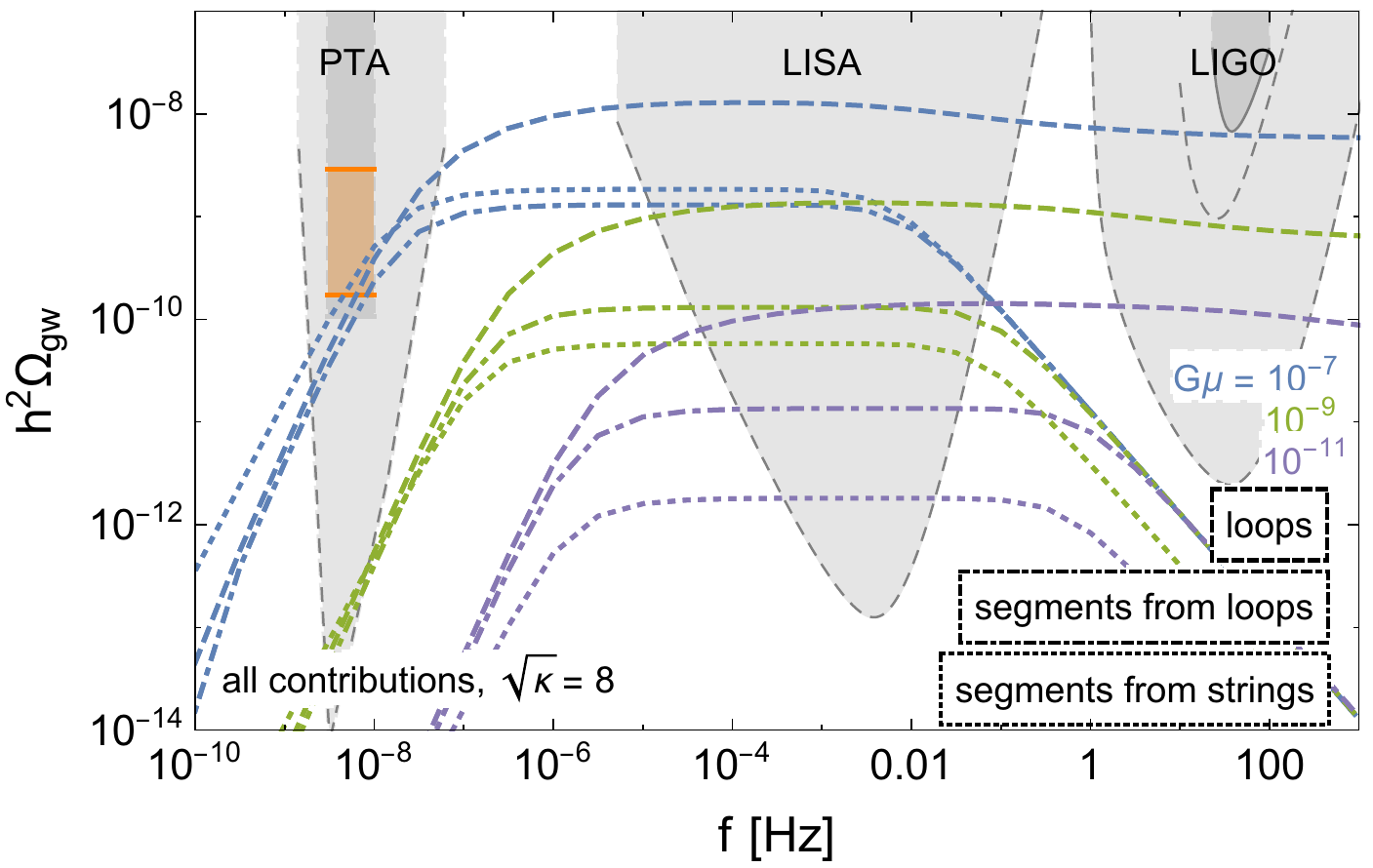} \hfill
\includegraphics[width = 0.48 \textwidth]{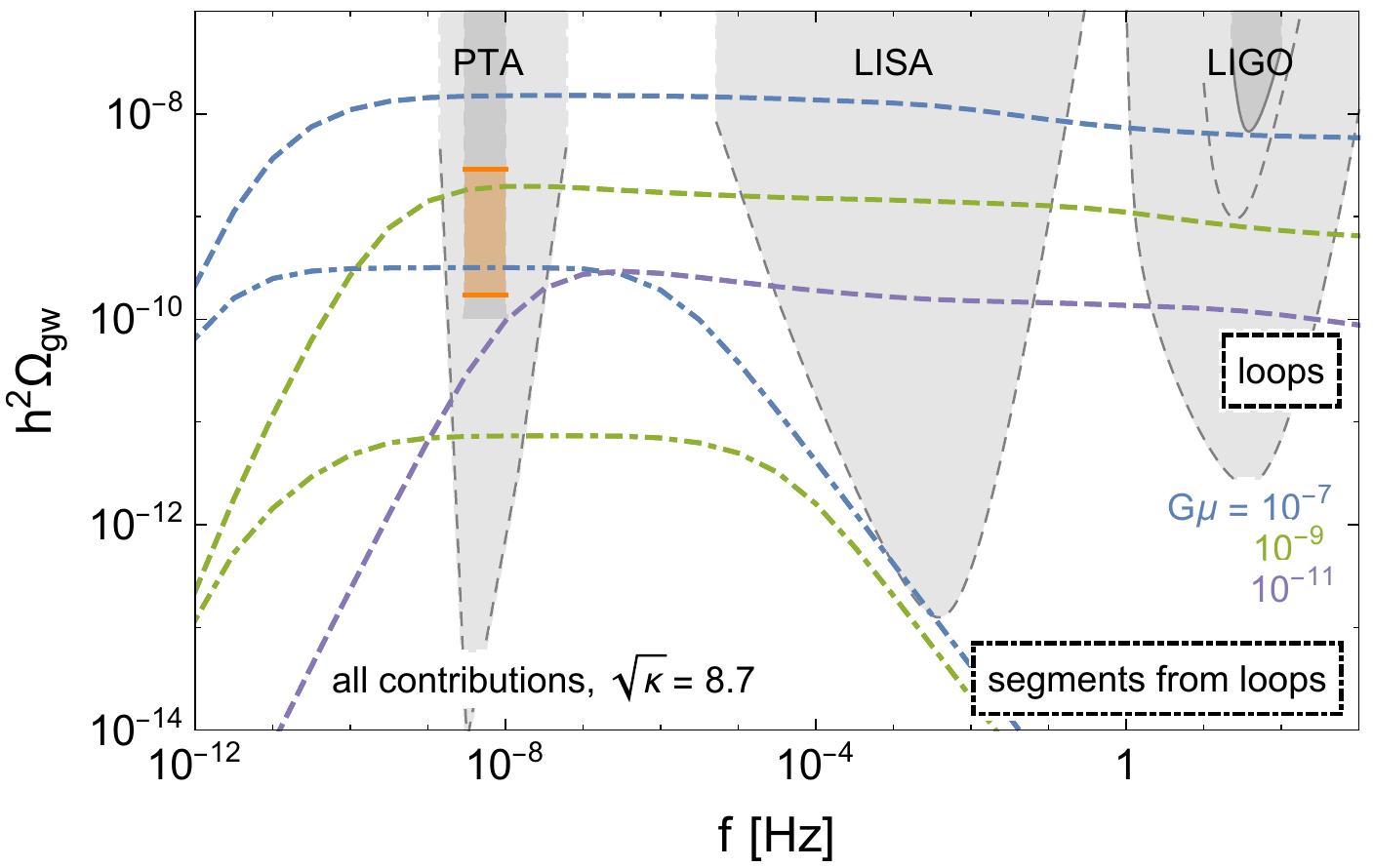}
\caption{Contributions to the GW spectrum in radiation domination for GWs sourced during the radiation era (left) and including contributions from the matter era (right). In the right panel the GW contribution from segments originating from long strings lies below the depicted plot range.}
\label{fig:contributions}
\end{figure}

\begin{figure}
\centering
\includegraphics[width = 0.48 \textwidth]{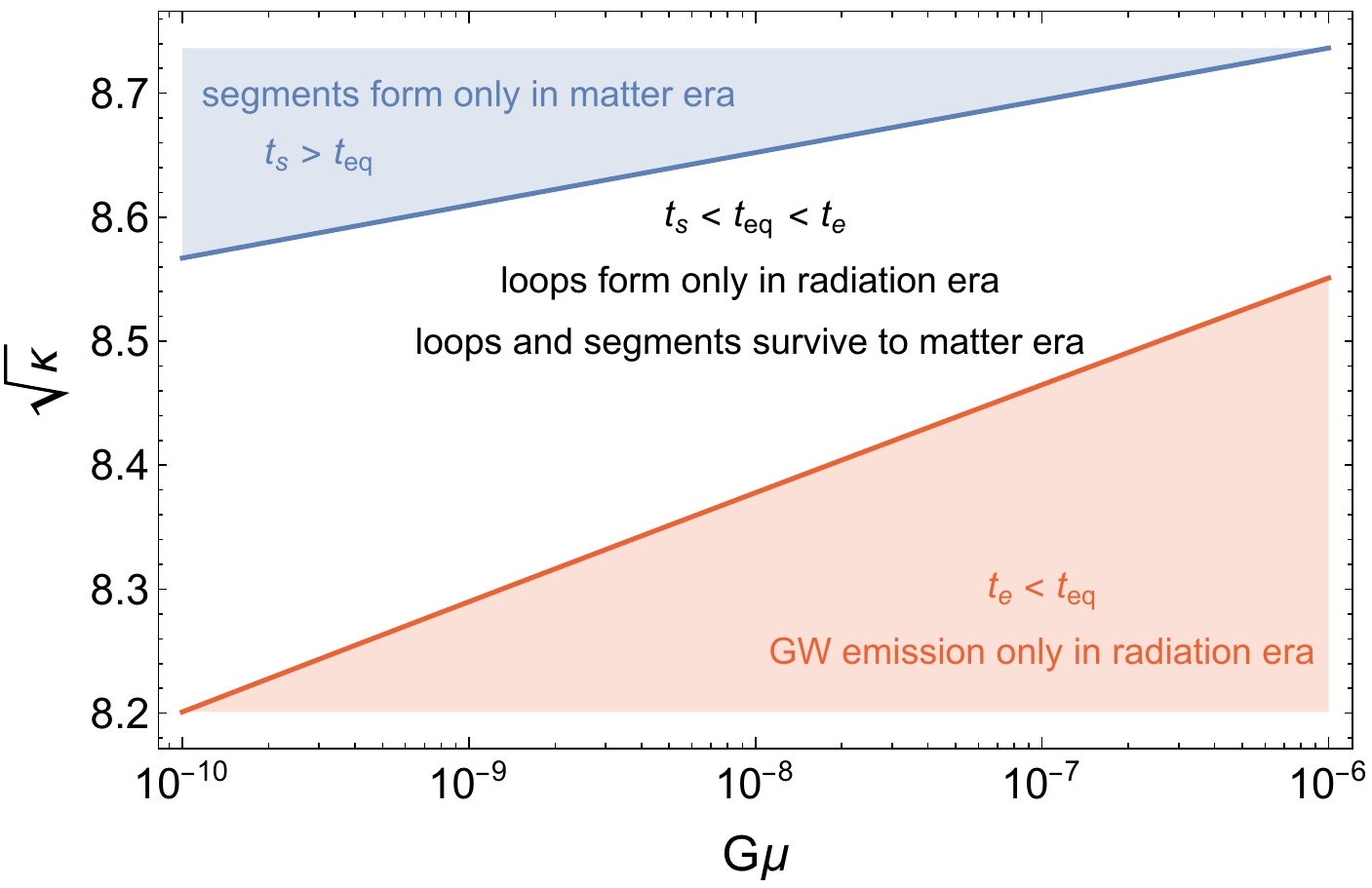} \hfill
\includegraphics[width = 0.48 \textwidth]{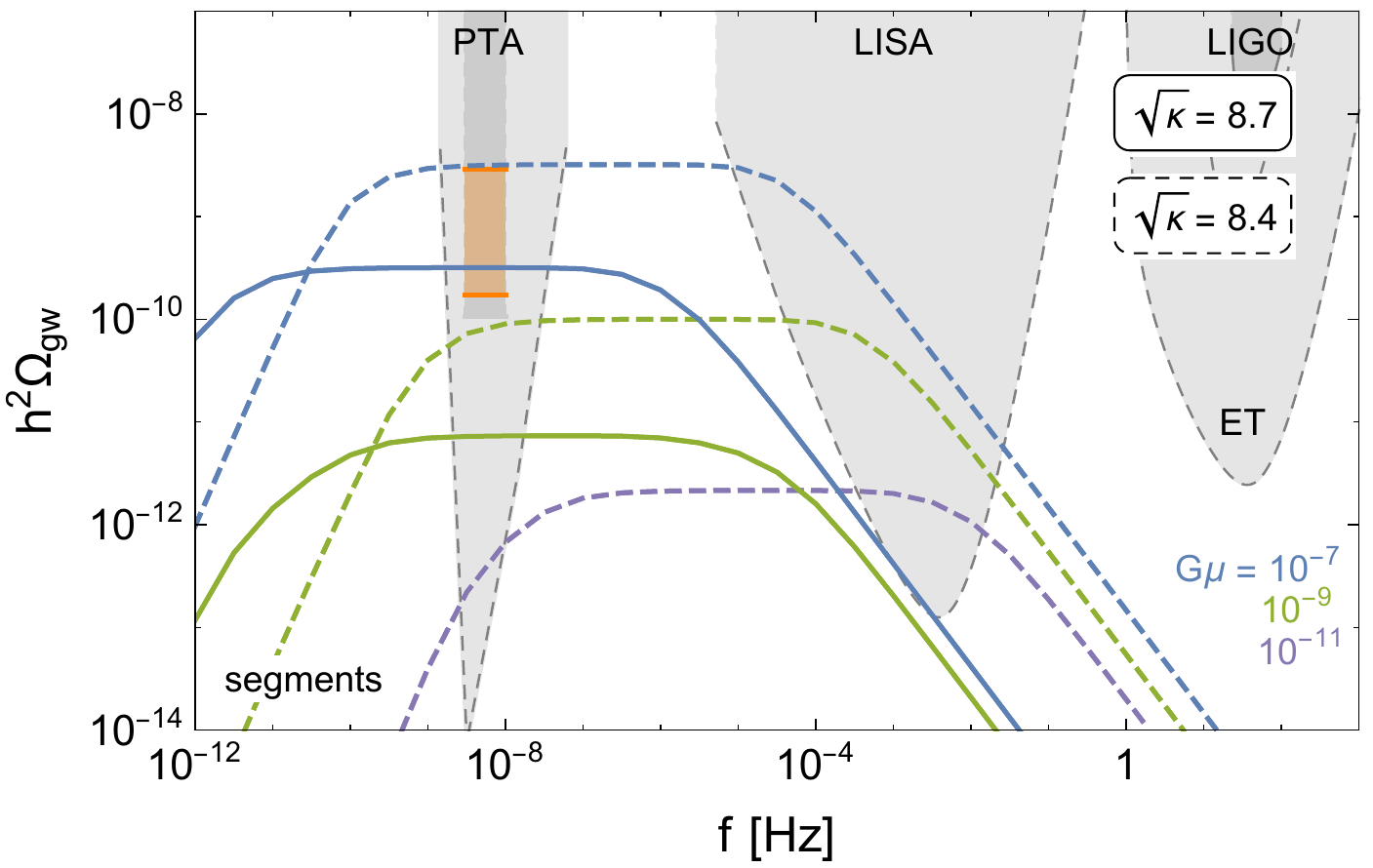}
\caption{GW spectrum from segments for long-lived strings. Left panel: Contributions from radiation and matter era in different parts of the model parameter space. Right panel: For large value of $\kappa$, the GW emission is restricted to the matter era and the amplitude of the spectrum is suppressed.}
\label{fig:segments}
\end{figure}

This subsection is dedicated to the derivation of the loop number density of metastable cosmic strings throughout different cosmological epochs. We begin with a detailed derivation of the loop number density during radiation domination, which underlies the analytical estimates presented in the main body of the text.
We then proceed to include the matter era, which we include in our numerical computations of the GW spectra. We follow a similar procedure for the segment number density in Secs.~\ref{subsec:segseg} and \ref{subsec:segloop}.
For a comparison of the various contributions both in radiation and matter domination, see Figs.~\ref{fig:contributions} and \ref{fig:segments}.
The left panel of Fig.~\ref{fig:segments}, moreover, illustrates the relevant time scales.

%%%%%%%%%%%%%%%%%%%%%%%%%%%%%%%%%%%%%%%%%%%%%%%%%%%%%%%%%%%%%%%%%%%%%%%%%%%%%%%%%%%%%%%%%%%%%%%%%%%%%

\subsubsection*{Radiation era}

%%%%%%%%%%%%%%%%%%%%%%%%%%%%%%%%%%%%%%%%%%%%%%%%%%%%%%%%%%%%%%%%%%%%%%%%%%%%%%%%%%%%%%%%%%%%%%%%%%%%%

The source term for loops is provided by the loop production function $f\left(\ell,t\right)$, and the time derivative of the loop length is controlled by the energy loss due to GW emission,
\begin{equation}
u\left(\ell,t\right) = - \Gamma G\mu \,,
\end{equation}
which yields the time-dependent length
\begin{equation}
\label{barl}
\bar{\ell}\left(t'\right) = \ell + \Gamma G\mu \left(t-t'\right) \,.
\end{equation}
For vanishing initial loop density $\overset{\circ}{n}_i$, one obtains in the scaling regime, $t < t_s$,
\begin{align}
\label{nloop}
\overset{\circ}{n}_<\left(\ell,t\right) = \int_{t_i}^t dt' \left(\frac{a\left(t'\right)}{a\left(t\right)}\right)^3 f\left(\bar{\ell},t'\right) e^{-\Gamma_d \left[\ell\left(t-t'\right) + \sfrac{1}{2}\,\Gamma G\mu\left(t-t'\right)^2\right]} \,.
\end{align}
For the BOS model, the loop production function is approximately given by
\begin{equation}
\label{fBOS}
f\left(\ell,t\right) = \frac{B}{\alpha^{3/2}\,t^4}\,\delta(\ell - \alpha t) \,,
\end{equation}
with $\Gamma G\mu \ll \alpha = 0.1$.
Hence, the density $\overset{\circ}{n}_<\left(\ell,t\right)$ of loops with length $\ell$ at time $t$ is determined by the number of loops that are produced at time $t' = \left(\ell +\Gamma G\mu\,t\right)/\alpha$ with size $\alpha\,t' = \ell + \Gamma G\mu\,t$.
Inserting Eqs.~\eqref{fBOS} and \eqref{barl} into Eq.~\eqref{solution} and setting $t_i = 0$, one obtains the loop number density in the radiation era,
\begin{equation}
\label{dbos0}
\overset{\circ}{n}_<\left(\ell,t\right) = \frac{B}{t^{3/2}\left(\ell + \Gamma G\mu t\right)^{5/2}}\,e^{-\Gamma_d \left[\ell\left(t-\ell/\alpha\right) + \sfrac{1}{2}\,\Gamma G\mu\left(t-\ell/\alpha\right)^2\right]} \,\Theta(\alpha t- l) \,.
\end{equation}
Since $t < t_s = 1/\Gamma_d^{1/2}$ and $\ell \leq \alpha\,t$, the two exponential damping terms are not effective.
The case of stable loops is obtained for $\Gamma_d \rightarrow 0$, i.e., $t_s \rightarrow \infty$.
In this limit, one obtains the loop number density \eqref{bos} of the BOS model,
\begin{equation}
\overset{\circ}{n}_<\left(\ell,t\right) \underset{t_s \rightarrow \infty}{\longrightarrow} \overset{\circ}{n}\left(\ell,t\right) = \frac{B}{t^{3/2}\left(\ell + \Gamma G\mu t\right)^{5/2}}\,\Theta\left(\alpha t- \ell\right) \,.
\end{equation}

%%%%%%%%%%%%%%%%%%%%%%%%%%%%%%%%%%%%%%%%%%%%%%%%%%%%%%%%%%%%%%%%%%%%%%%%%%%%%%%%%%%%%%%%%%%%%%%%%%%%%

After the initial scaling regime, for $t > t_s= 1/\Gamma_d^{1/2}$, one has to use Eq.~\eqref{solution} with vanishing loop production function and an initial number density determined by the matching condition
\begin{equation}
\label{matching}
\overset{\circ}{n}_>\left(\bar{\ell}\left(t_s\right),t_s\right) = \overset{\circ}{n}_<\left(\bar{\ell}\left(t_s\right),t_s\right) \simeq 
\frac{B}{t_s^{3/2}\left(\bar{\ell}\left(t_s\right) + \Gamma G\mu t_s\right)^{5/2}} \,\Theta\left(\alpha t_s- \bar{\ell}\left(t_s\right)\right) \,.
\end{equation}
Using $\bar{\ell}\left(t_s\right) + \Gamma G\mu\,t_s = \ell + \Gamma G\mu\,t$ and Eqs.~\eqref{solution} and \eqref{matching}, one obtains
\begin{equation}
\label{dbos1}
\overset{\circ}{n}_>\left(\ell,t\right) = \frac{B}{t^{3/2}\left(\ell + \Gamma G\mu t\right)^{5/2}}\,e^{-\Gamma_d\left[\ell\left(t-t_s\right) + \sfrac{1}{2}\,\Gamma G\mu\left(t-t_s\right)^2 \right]} \,\Theta\left(\alpha t_s- \bar{\ell}\left(t_s\right)\right) \, \Theta\left(t_{\rm{eq}}-t\right)\,, 
\end{equation}
The result differs from the number density of stable loops in Eq.~\eqref{bos} by two damping terms, which become relevant for $\ell t \gtrsim 1/\Gamma_d$ and $t^2 \gtrsim 2/\left(\Gamma_d \Gamma G\mu\right) = t_e^2$, respectively.
A further important difference is the argument of the Heaviside theta functions.
Since $t_s < t$ and $\bar{\ell}\left(t_s\right) = \ell + \Gamma G\mu \left(t-t_s\right) > \ell$, the constraint for $\overset{\circ}{n}_>$ is more stringent, which reflects the fact that only loops produced before $t_s$ contribute to the number density.

%%%%%%%%%%%%%%%%%%%%%%%%%%%%%%%%%%%%%%%%%%%%%%%%%%%%%%%%%%%%%%%%%%%%%%%%%%%%%%%%%%%%%%%%%%%%%%%%%%%%%

\subsubsection*{Matter era}

%%%%%%%%%%%%%%%%%%%%%%%%%%%%%%%%%%%%%%%%%%%%%%%%%%%%%%%%%%%%%%%%%%%%%%%%%%%%%%%%%%%%%%%%%%%%%%%%%%%%%

After matter--radiation equality, $t > t_{\rm eq}$, we distinguish between loops produced in the radiation era (but surviving until the matter era) and loops formed in the matter era.
For $ t < t_s$, the loop number densities are given by the corresponding expressions found for topologically stable cosmic strings in the BOS model~\cite{Blanco-Pillado:2013qja},
\begin{align}
\label{eq:bos_rm}
\overset{\circ}{n}^{\text{rm}}_<\left(\ell,t\right) &  = \frac{B}{\left(\ell+\Gamma G\mu t\right)^{5/2}}\frac{t_{\rm eq}^{1/2}}{t^2}\,\Theta\left(\alpha t - \ell\right) \,,\\
\overset{\circ}{n}^\text{m}_<\left(\ell,t\right) & = \frac{A_1 - A_2\left(\ell/t\right)^\beta}{t^2 \left(\ell + \Gamma G \mu t\right)^2}\,\Theta\left(\gamma t - \ell\right) \Theta\left(t_s - t_{\rm eq}\right) \,,
\end{align}
with $A_1 = 0.27$, $A_2 = 0.45$, $\beta = 0.31$, and $\gamma = 0.18$.
Here, the second theta function in the expression for $\overset{\circ}{n}^{\rm m}_<$ reflects the fact that loop production only occurs at $t < t_s$.
To obtain the number density at $t > t_s$, we use Eq.~\eqref{solution} with a vanishing loop production function and initial conditions determined by the matching condition at $t = t_s$.
Analogously to Eq.~\eqref{dbos1}, this yields for the loop number densities at $t > t_s,t_\text{eq}$,
\begin{align}
\overset{\circ}{n}^{\rm rm}_>\left(\ell,t\right) &  = \frac{B}{\left(\ell+\Gamma G\mu t\right)^{5/2}}\frac{t_{\rm eq}^{1/2}}{t^2} \,e^{-\Gamma_d \left[\ell\left(t-t_s\right) + \sfrac{1}{2}\,\Gamma G\mu\left(t-t_s\right)^2\right]}\,\Theta\left(\alpha t_s- \bar{\ell}\left(t_s\right)\right) \,, \nonumber \\
\overset{\circ}{n}^{\rm m}_>\left(\ell,t\right) & = \frac{A_1 - A_2\left(\ell/t\right)^\beta}{t^2 \left(\ell + \Gamma G \mu t\right)^2}\,e^{-\Gamma_d \left[\ell\left(t-t_s\right) + \sfrac{1}{2}\,\Gamma G\mu\left(t-t_s\right)^2\right]}\,\Theta\left(t_s - t_{\rm eq}\right) \Theta\left(\gamma t_s- \bar{\ell}\left(t_s\right)\right) \label{eq:nmg} \,,
\end{align}
where $t_s$ can be either before or after $t_{\rm eq}$.
Eqs.~\eqref{eq:bos_rm} to \eqref{eq:nmg}, together with Eqs.~\eqref{bos} and \eqref{dbos1}, describe the loop number density of metastable cosmic strings throughout cosmic history.

%%%%%%%%%%%%%%%%%%%%%%%%%%%%%%%%%%%%%%%%%%%%%%%%%%%%%%%%%%%%%%%%%%%%%%%%%%%%%%%%%%%%%%%%%%%%%%%%%%%%%

\subsection{Segments sourced by long strings} 
\label{subsec:segseg}

%%%%%%%%%%%%%%%%%%%%%%%%%%%%%%%%%%%%%%%%%%%%%%%%%%%%%%%%%%%%%%%%%%%%%%%%%%%%%%%%%%%%%%%%%%%%%%%%%%%%%

%%%%%%%%%%%%%%%%%%%%%%%%%%%%%%%%%%%%%%%%%%%%%%%%%%%%%%%%%%%%%%%%%%%%%%%%%%%%%%%%%%%%%%%%%%%%%%%%%%%%%

For string segments from long strings, whose number density we denote by $\tilde{n}^{(s)}\left(\ell,t\right)$, the source term is the splitting of one segment (or long string) into two segments~\cite{Leblond:2009fq},
\begin{equation}
\label{Stilden}
S\left(\ell,t\right) = 2\,\Gamma_d\,\int_\ell^\infty d\ell'\,\tilde{n}^{(s)}\left(\ell',t\right) \,.
\end{equation}
Long superhorizon string segments gain length by Hubble stretching and shrink due to energy loss by loop production.
To compute this effect directly is difficult, but since $u\left(\ell,t\right) \propto \ell$, this function can be determined by demanding that at early times, $t < t_s$, the long segments exhibit scaling, i.e., $\mu \int_\ell d\ell\,\ell\,\tilde{n}\left(\ell,t\right) \sim \mu/t^2$.
One then finds \cite{Leblond:2009fq},
\begin{equation}
\label{utilden}
u\left(\ell,t\right) = 3H\left(t\right)\ell - \frac{2\ell}{t} \quad\Rightarrow\quad \bar{\ell}\left(t'\right) =\left(\frac{a\left(t'\right)}{a\left(t\right)}\right)^3\left(\frac{t}{t'}\right)^2 \ell \,.
\end{equation} 

%%%%%%%%%%%%%%%%%%%%%%%%%%%%%%%%%%%%%%%%%%%%%%%%%%%%%%%%%%%%%%%%%%%%%%%%%%%%%%%%%%%%%%%%%%%%%%%%%%%%%

\subsubsection*{Radiation era}

%%%%%%%%%%%%%%%%%%%%%%%%%%%%%%%%%%%%%%%%%%%%%%%%%%%%%%%%%%%%%%%%%%%%%%%%%%%%%%%%%%%%%%%%%%%%%%%%%%%%%

Using \eqref{Stilden} and \eqref{utilden}, Eq.~\eqref{generalkinetic} in the radiation era has a simple solution ($t < t_s$),
\begin{equation}
\label{stilden}
\tilde{n}^{(s)}_<\left(\ell,t\right) = C\,\Gamma_d^2\,e^{-2\,\Gamma_d\,\ell\,t} \,,
\end{equation}
Since the kinetic equation for $\tilde{n}^{(s)}_<$ is homogeneous, the normalization is not fixed and $C$ is a free parameter that can be determined using the matching condition~\eqref{matchrho}, yielding $C = 4/\xi^2$.

%%%%%%%%%%%%%%%%%%%%%%%%%%%%%%%%%%%%%%%%%%%%%%%%%%%%%%%%%%%%%%%%%%%%%%%%%%%%%%%%%%%%%%%%%%%%%%%%%%%%%

After the initial scaling regime time, for $t > t_s$, typical segments enter the horizon.
They split into smaller segments and shrink due to gravitational radiation, i.e.,
\begin{equation}
S\left(\ell,t\right) = 2\,\Gamma_d\,\int_\ell^\infty d\ell'\,\tilde{n}^{(s)}\left(\ell',t\right) \,, \quad u\left(\ell,t\right) = - \tilde{\Gamma}G\mu \,.
\end{equation}
The solution of the corresponding kinetic equation~\eqref{generalkinetic}, which matches with $\tilde{n}^{(s)}_<\left(\ell,t\right)$ at $t_s$,
\begin{equation}
\label{tildematch}
\tilde{n}^{(s)}_>\left(\bar{\ell}\left(t_s\right),t_s\right) = \tilde{n}^{(s)}_<\left(\bar{\ell}\left(t_s\right),t_s\right) \,,
\end{equation}
is known analytically \cite{Leblond:2009fq} and given by Eq.~\eqref{ntilde},
\begin{equation}
\tilde{n}^{(s)}_>\left(\ell,t\right) = C\,\frac{\Gamma_d^2}{4} \frac{\left(t+t_s\right)^2}{\sqrt{t^3t_s}} e^{-\Gamma_d\left[\ell\left(t+t_s\right) + \frac{1}{2} \tilde{\Gamma} G\mu \left(t-t_s\right)\left(t+3t_s\right)\right]} \,\Theta\left(t_{\rm eq}-t\right) \,.
\end{equation}
The two damping terms are essentially the same as in Eq.~\eqref{dbos}, but there is no restriction on the segment lengths.
The segments have decayed after $\tilde{t}_e = \big(\tilde{\Gamma} G\mu\big)^{-1/2}\,t_s$.
This can be immediately generalized to include later times $t > t_{\rm eq}$, as long as the matching time remains in radiation domination, $t_s < t_{\rm eq}$.
In this case, we obtain
\begin{equation}
\tilde{n}_>^{\textrm{rm}\,(s)}\left(\ell,t\right) = C\,\frac{\Gamma_d^2}{4}\left(\frac{t_{\rm eq}}{t}\right)^2 \frac{\left(t+t_s\right)^2}{\sqrt{t_{\rm eq}^3t_s}}\,e^{-\Gamma_d \left[\ell\left(t+t_s\right) + \sfrac{1}{2}\,\tilde{\Gamma} G\mu\left(t-t_s\right)\left(t+3t_s\right)\right]}\,\Theta\left(t-t_{\rm eq}\right) \,.
\end{equation}

%%%%%%%%%%%%%%%%%%%%%%%%%%%%%%%%%%%%%%%%%%%%%%%%%%%%%%%%%%%%%%%%%%%%%%%%%%%%%%%%%%%%%%%%%%%%%%%%%%%%%

%%%%%%%%%%%%%%%%%%%%%%%%%%%%%%%%%%%%%%%%%%%%%%%%%%%%%%%%%%%%%%%%%%%%%%%%%%%%%%%%%%%%%%%%%%%%%%%%%%%%%

\subsubsection*{Matter era}

%%%%%%%%%%%%%%%%%%%%%%%%%%%%%%%%%%%%%%%%%%%%%%%%%%%%%%%%%%%%%%%%%%%%%%%%%%%%%%%%%%%%%%%%%%%%%%%%%%%%%

We proceed to consider the remaining case of $t_s > t_{\rm eq}$.
For $t_{\rm eq} < t < t_s$, solving the kinetic equation with~\eqref{Stilden} and \eqref{utilden} (where for matter domination $u\left(\ell,t\right) = 0$) yields
\begin{equation}
\tilde n_<^{\textrm{m}\,(s)}\left(\ell,t\right) = \frac{\Gamma_d^2}{\xi_m^2}\, e^{-\Gamma_d\,\ell\,t} \,,
\end{equation}
where the prefactor has been determined by matching at $t = t_s$ to the scaling regime of topologically stable cosmic strings in matter domination, $\rho_\infty\left(t\right) = \mu^2/\left(t^2 \xi_m^2\right)$ with $\xi_m = 0.625$~\cite{Auclair:2019wcv}.

%%%%%%%%%%%%%%%%%%%%%%%%%%%%%%%%%%%%%%%%%%%%%%%%%%%%%%%%%%%%%%%%%%%%%%%%%%%%%%%%%%%%%%%%%%%%%%%%%%%%%

For $t > t_s > t_{\rm eq}$, inserting the ansatz
\begin{equation}
\label{ansatz}
\tilde{n}_>^{\textrm{m}\,(s)}\left(\ell,t\right) = C_s\left(t\right)\,e^{-\Gamma_d\,\ell\,t} 
\end{equation}
into the kinetic equation~\eqref{generalkinetic} yields
\begin{equation}
\label{eq:Cs} 
\dot C_s\left(t\right) = - \Gamma_d\,\tilde{\Gamma} G \mu\,t\,C_s\left(t\right)  \,,\quad C_s\left(t_s\right) = \frac{\Gamma_d^2}{\xi_m^2} \,, 
\end{equation}
which is solved (for $\tilde \Gamma = \Gamma$) by
\begin{equation}
C_s\left(t\right) = \frac{1}{t_s^4\,\xi_m^2}\,e^{-\sfrac{1}{2}\,\Gamma_d\,\Gamma G\mu\left(t^2-t_s^2\right)} \,,
\end{equation}
yielding for the number density
\begin{equation}
\label{ntilde_m_s}
\tilde{n}_>^{\textrm{m}\,(s)}\left(\ell,t\right) =  \frac{1}{t_s^4\,\xi_m^2}\,e^{-\Gamma_d \left[\ell\,t + \sfrac{1}{2}\,\Gamma G\mu\left(t^2-t_s^2\right)\right]} \,.
\end{equation}

%%%%%%%%%%%%%%%%%%%%%%%%%%%%%%%%%%%%%%%%%%%%%%%%%%%%%%%%%%%%%%%%%%%%%%%%%%%%%%%%%%%%%%%%%%%%%%%%%%%%%

\subsection{Segments sourced by loops}
\label{subsec:segloop}

%%%%%%%%%%%%%%%%%%%%%%%%%%%%%%%%%%%%%%%%%%%%%%%%%%%%%%%%%%%%%%%%%%%%%%%%%%%%%%%%%%%%%%%%%%%%%%%%%%%%%

In addition to decaying long strings, decaying loops also yield string segments, whose number density we will denote by $\tilde{n}^{(l)}\left(\ell,t\right)$.
In the absence of simulations and analytical calculations, we treat them in the same way as segments from long strings. 
The kinetic equation for $\tilde{n}^{(l)}\left(\ell,t\right)$ at $t > t_s$ then becomes a linear partial integro-differential equation,
\begin{equation}
\label{kintildefull}
\partial_t\, \tilde{n}^{(l)}_>\left(\ell,t\right) = - \left[3H\left(t\right) + \Gamma_d\,\ell - \tilde{\Gamma}\,G\mu\,\partial_\ell \right] \tilde{n}^{(l)}_>\left(\ell,t\right) + 2\,\Gamma_d\,\int_\ell^{\infty}d\ell'\,\tilde{n}^{(l)}_>\left(\ell',t\right) + \Gamma_d\,\ell\,\overset{\circ}{n}_>\left(\ell,t\right) \,,
\end{equation}
where loop decays now act as an additional source term.
For $\tilde{\Gamma} = \Gamma$, this equation can be formally solved by an infinite series.
To see this, we can first write the solution of Eq.~\eqref{kintildefull} in exactly the same way as the general solution in Eq.~\eqref{solution}, with the source term inside the $t'$ integral given by
\begin{equation}
\label{eq:source}
S\left(\bar{\ell}\left(t'\right),t'\right) =  2\,\Gamma_d\,\int_{\bar{\ell}\left(t'\right)}^{\infty}d\ell'\,\tilde{n}^{(l)}_>\left(\ell',t'\right) + \Gamma_d\,\bar{\ell}\left(t'\right)\,\overset{\circ}{n}_>\left(\bar{\ell}\left(t'\right),t'\right) \,.
\end{equation}
The expression for $\tilde{n}^{(l)}_>\left(\ell,t\right)$ thus obtained refers to itself, via the $\tilde{n}^{(l)}_>\left(\ell',t'\right)$ term inside the $\ell'$ integral.
This dependence can be removed by an iterative procedure.
In the next step, we take the full expression for $\tilde{n}^{(l)}_>\left(\ell,t\right)$ that we just derived and insert it back into itself, more precisely, into the $\ell'$ integral in the source term.
Repeating this step over and over again then results in an infinite series for $\tilde{n}^{(l)}_>\left(\ell,t\right)$, which no longer refers back to itself and which can hence be systematically evaluated order by order without any prior knowledge of the final solution,
\begin{equation}
\label{eq:series}
\tilde{n}_>^{(l)}\left(\ell,t\right) = \sum_{i=1}^\infty \tilde{n}_>^{(l,i)}\left(\ell,t\right) \,,\quad \tilde{n}_>^{(l,i+1)}\left(\ell,t\right) = 2\,\Gamma_d \int_{t_i}^t dt' \int_{\bar{\ell}\left(t'\right)}^\infty d\ell' A\left(\ell,t,t'\right)\tilde{n}_>^{(l,i)}\left(\ell',t'\right) \,.
\end{equation}
Here, the function $A$ accounts for the cosmological redshift, decay into smaller segments, and GW emission of string segments with length $\ell$ at time $t$ in the time interval from $t'$ to $t$,
\begin{equation}
A\left(\ell,t,t'\right) = \left(\frac{a\left(t'\right)}{a\left(t\right)}\right)^3\,e^{-\Gamma_d \left[\ell\left(t-t'\right) + \sfrac{1}{2}\,\tilde{\Gamma} G\mu\left(t-t'\right)^2\right]} \,.
\end{equation}
The interpretation of the infinite series in Eq.~\eqref{eq:series} is straightforward.
The first term in the series, $\tilde{n}_>^{(l,1)}\left(\ell,t\right)$, describes the first generation of segments from loops, i.e., segments that form in consequence of monopole pair creation events on string loops; the second term, $\tilde{n}_>^{(l,2)}\left(\ell,t\right)$, describes the second generation of segments from loops, i.e., segments that form in consequence of monopole pair creation events on first-generation segments; and so on and so forth.

%%%%%%%%%%%%%%%%%%%%%%%%%%%%%%%%%%%%%%%%%%%%%%%%%%%%%%%%%%%%%%%%%%%%%%%%%%%%%%%%%%%%%%%%%%%%%%%%%%%%%

In order to evaluate the infinite series term by term, following the iterative procedure described in Eq.~\eqref{eq:series}, one needs to know the first term $\tilde{n}_>^{(l,1)}\left(\ell,t\right)$.
This term simply follows the general solution in Eq.~\eqref{solution} and the source term in Eq.~\eqref{eq:source} after dropping the $\ell'$ integral contribution in Eq.~\eqref{eq:source}.
In other words, it follows from the kinetic equation in Eq.~\eqref{kintildefull} after omitting the source term that describes the decay of string segments into smaller segments,
\begin{equation}
\label{kintildefull1st}
\partial_t\, \tilde{n}^{(l,1)}_>\left(\ell,t\right) = - \left[3H\left(t\right) + \Gamma_d\,\ell - \tilde{\Gamma}\,G\mu\,\partial_\ell \right] \tilde{n}^{(l,1)}_>\left(\ell,t\right) + \Gamma_d\,\ell\,\overset{\circ}{n}_>\left(\ell,t\right) \,.
\end{equation}
In contrast to Eq.~\eqref{kintildefull}, this is again a partial differential equation (i.e., no longer a partial integro-differential equation), which can be solved using the standard methods outlined at the beginning of this section. 
During the radiation era and making use of Eq.~\eqref{solution}, we thus obtain
\begin{equation}
\tilde{n}_>^{(l,1)}\left(\ell,t\right) = \Gamma_d\left[\ell\left(t-t_s\right) + \frac{1}{2}\,\Gamma G\mu\left(t-t_s\right)^2\right] \overset{\circ}{n}_>\left(\ell,t\right) \,,
\end{equation}
where we have used that $\tilde{n}_>^{(l,1)}\left(\ell,t\right) = 0$ at $t=t_s$.
Similarly, we obtain during the matter era
\begin{equation}
\label{eq:nfirstRD}
\tilde{n}_>^{\textrm{m}\,(l,1)}\left(\ell,t\right) = \tilde{n}_>^{\textrm{rm}\,(l,1)}\left(\ell,t\right) + \tilde{n}_>^{\textrm{mm}\,(l,1)}\left(\ell,t\right)
\end{equation}
where the segment densities $\tilde{n}_>^{\textrm{rm}\,(l,1)}$ and $\tilde{n}_>^{\textrm{mm}\,(l,1)}$ are induced by loops that are produced during radiation and matter domination, respectively, and all of which decay during matter domination.
The first term, $\tilde{n}_>^{\textrm{rm}\,(l,1)}$, corresponds to the straightforward continuation of $\tilde{n}_>^{(l,1)}$ in Eq.~\eqref{eq:nfirstRD},
\begin{equation}
\tilde{n}_>^{\textrm{rm}\,(l,1)}\left(\ell,t\right) = \Gamma_d\left[\ell\left(t-t_s\right) + \frac{1}{2}\,\Gamma G\mu\left(t-t_s\right)^2\right] \overset{\circ}{n}_>^{\rm{rm}}\left(\ell,t\right)
\end{equation}
while the second term, $\tilde{n}_>^{\textrm{mm}\,(l,1)}$, turns out to be less compact because of the slightly more complicated form of the loop number density during matter domination, $\overset{\circ}{n}^{\rm m}_>$, in Eq.~\eqref{eq:nmg},
\begin{align}
& \tilde{n}_>^{\textrm{mm}\,(l,1)}\left(\ell,t\right) = \frac{\Gamma_d}{t^2\left(\ell + \Gamma G\mu\,t\right)^2}\,e^{-\Gamma_d\left[\ell\left(t-t_s\right) + \sfrac{1}{2}\,\Gamma G\mu\left(t-t_s\right)^2\right]}\:\bigg\{A_1\left[\ell\left(t-t_s\right) + \frac{1}{2}\,\Gamma G\mu\left(t-t_s\right)^2\right] \nonumber\\
& + A_2\left(\ell + \Gamma G\mu\,t\right)^{1+\beta} \left[F_2\left(t\right)-F_1\left(t\right)-F_2\left(t_s\right)+F_1\left(t_s\right)\right] \bigg\} \:\Theta\left(\gamma\,t_s - \bar{\ell}\left(t_s\right)\right)\Theta\left(t_s-t_{\rm eq}\right)\,,
\end{align}
where the auxiliary functions $F_1$ and $F_2$ are given in terms of the hypergeometric function ${}_2F_1$,
\begin{equation}
F_n\left(x\right) = {}_2F_1\left(n-\beta,-\beta;n+1-\beta;\frac{\Gamma G\mu}{\ell + \Gamma G\mu\,t}\,x\right)\left(\frac{\Gamma G\mu}{\ell + \Gamma G\mu\,t}\right)^{n-1} \frac{x^{n-\beta}}{n-\beta} \,.
\end{equation}

%%%%%%%%%%%%%%%%%%%%%%%%%%%%%%%%%%%%%%%%%%%%%%%%%%%%%%%%%%%%%%%%%%%%%%%%%%%%%%%%%%%%%%%%%%%%%%%%%%%%%

The above expressions for the first term in the infinite series are the starting point for numerically evaluating the higher terms in the series.
We perform such a numerical analysis, which reveals that the series rapidly converges after the first few terms. 
Moreover, we find that the GW spectrum computed based on the full result for $\tilde{n}^{(l)}$ is well approximated by the GW spectrum computed based on the first term, $\tilde{n}^{(l,1)}$, times a numerical fudge factor $\sigma$.
For all practical purposes in this paper, this observation allows us to replace to the full result for $\tilde{n}^{(l)}$ by $\tilde{n}^{(l,1)}$ times the fudge factor $\sigma$, even though the functional dependence of $\tilde{n}^{(l)}$ and $\tilde{n}^{(l,1)}$ on $\ell$ and $t$ is not identical,
\begin{equation}
\label{eq:nSL_r}
\tilde{n}^{(l)}\left(\ell,t\right) \rightarrow \sigma\, \tilde{n}^{(l,1)}\left(\ell,t\right) \,, \quad \sigma \simeq 5 \,,
\end{equation}
and 
\begin{equation}
\label{eq:nSL_m}
\tilde{n}_>^{\textrm{m}\,(l)}\left(\ell,t\right) \rightarrow \sigma^{\rm m}\left[\tilde{n}_>^{\textrm{rm}\,(l,1)}\left(\ell,t\right) + \tilde{n}_>^{\textrm{mm}\,(l,1)}\left(\ell,t\right)\right]
\end{equation}
with $\sigma^{\rm m} \simeq \sigma \simeq 5$ for $t_s < t_{\rm eq}$ and 
\begin{align}
\sigma^{\rm m} \simeq
\begin{cases}
 1  & \text{for~} G\mu < 10^{-9.5}                                \\
 5  & \text{for~} G\mu > 10^{-9.5} \text{~and loops formed in RD} \\
15  & \text{for~} G\mu > 10^{-9.5} \text{~and loops formed in MD}
\end{cases}
\end{align}
for $t_s > t_\text{eq}$.

%%%%%%%%%%%%%%%%%%%%%%%%%%%%%%%%%%%%%%%%%%%%%%%%%%%%%%%%%%%%%%%%%%%%%%%%%%%%%%%%%%%%%%%%%%%%%%%%%%%%

\bibliographystyle{JHEP}
\bibliography{arxiv_2}

%%%%%%%%%%%%%%%%%%%%%%%%%%%%%%%%%%%%%%%%%%%%%%%%%%%%%%%%%%%%%%%%%%%%%%%%%%%%%%%%%%%%%%%%%%%%%%%%%%%%

\end{document}